\documentclass[journal]{IEEEtran}

\usepackage{booktabs}
\usepackage{tabularx}

\usepackage{amssymb}
\usepackage{amsbsy}
\usepackage{amsmath}
\usepackage{bm}
\usepackage{verbatim}
\usepackage{cite}
\usepackage{mathrsfs}
\usepackage{amsfonts}
\usepackage{graphicx}
\usepackage[tight,footnotesize]{subfigure}
\usepackage[10pt]{moresize}
\usepackage{array}
\usepackage{color}
\usepackage{epsfig}
\usepackage{stfloats}
\usepackage{balance}

\usepackage{hyperref}  

\usepackage{amssymb} 

\usepackage[noend]{algpseudocode}

\usepackage{algorithmicx,algorithm}

\usepackage{cite}

\usepackage{setspace}
\usepackage{cases}

\usepackage{graphicx}
\usepackage{epstopdf}
\usepackage{multirow}
\usepackage{extarrows}
\newcommand{\subparagraph}{}
\usepackage{titlesec}
\titlespacing{\section}{0pt}{2 ex plus .0ex minus .0ex}{1ex plus .0ex}
\titlespacing{\subsection}{0pt}{1.5 ex plus .0ex minus .0ex}{0.8 ex plus 0.0ex}
\titlespacing{\subsubsection}{0pt}{0.5ex plus .0ex minus .0ex}{0.0ex plus .0ex}

\usepackage{amsmath} 
\allowdisplaybreaks[4]

\ifCLASSINFOpdf

\else

\fi

\begin{document}

\title{An IRS Backscatter Enabled Integrated Sensing, Communication and Computation System}

\author{{Sai~Xu,~\IEEEmembership{Member,~IEEE,}
~Yanan Du, ~\IEEEmembership{Graduate Student Member,~IEEE,}
~Jiliang Zhang, ~\IEEEmembership{Senior~Member,~IEEE,}
~Jiangzhou Wang, ~\IEEEmembership{Fellow,~IEEE,}
and~Jie~Zhang,~\IEEEmembership{Senior~Member,~IEEE}
\thanks{%
Sai~Xu (e-mail: \texttt{xusai@nwpu.edu.cn}) is with the School of Cybersecurity, Northwestern Polytechnical University, Xi'an, Shaanxi, 710072, China, and also the Department of Electronic and Electrical Engineering, University of Sheffield, Sheffield, S1 4ET, UK. Yanan~Du (e-mail: \texttt{ynduyndu@163.com}) is with the School of Cybersecurity, Northwestern Polytechnical University, Xi'an, Shaanxi, 710072, China. Jiangzhou Wang (e-mail: \texttt{j.z.wang@kent.ac.uk}) is with the School of Engineering and Digital Arts, University of Kent, Canterbury, CT2 7NT, UK. Jiliang~Zhang (e-mail: \texttt{jiliang.zhang@sheffield.ac.uk}) and Jie~Zhang (e-mail: \texttt{jie.zhang@sheffield.ac.uk}) are with the Department of Electronic and Electrical Engineering, University of Sheffield, Sheffield, S1 4ET, UK.}}
\thanks{Manuscript received XX XX, XXXX; revised XX XX, XXXX. }}
\maketitle
\maketitle
%
%
%


%
%
%
%
\begin{abstract}
This paper proposes to leverage intelligent reflecting surface (IRS) backscatter to realize radio-frequency-chain-free uplink-transmissions (RFCF-UT). In this communication paradigm, IRS works as an information carrier, whose elements are capable of adjusting their amplitudes and phases to collaboratively portray an electromagnetic image like a dynamic quick response (QR) code, rather than a familiar reflection device, while a full-duplex base station (BS) is used as a scanner to collect and recognize the information on IRS. To elaborate it, an integrated sensing, communication and computation system as an example is presented, in which a dual-functional radar-communication BS simultaneously detects the target and collects the data from user equipments each connected to an IRS. Based on the established model, partial and binary data offloading strategies are respectively considered. By defining a performance metric named weighted throughput capacity (WTC), two maximization problems of WTC are formulated. According to the coupling degree of optimization variables in the objective function and the constraints, each optimization problem is firstly decomposed into two subproblems. Then, the methods of linear programming, fractional programming, integer programming and alternative optimization are developed to solve the subproblems. The simulation results demonstrate the achievable WTC of the considered system, thereby validating RFCF-UT.
\end{abstract}
%
\IEEEpeerreviewmaketitle
\section{Introduction}
\IEEEPARstart{L}IKE well-known computation offloading~\cite{Mach2017Mobile} and transmitting beamforming design~\cite{Agiwal2016Next} being both capable of lightening user equipments (UEs), it is of great importance to move transmitting radio-frequency (RF) chains from UEs to their associated base stations (BSs). That is primarily because BSs are generally more advanced with more antennas, higher transmitting power, more channel and signal knowledge, etc. For UE lightweight, this paper proposes a novel communication paradigm named radio-frequency-chain-free uplink-transmissions (RFCF-UT), whose key idea is to employ intelligent reflecting surface (IRS)~\cite{Wu2020Towards} to map the information to the complex reflection coefficients of its elements. To be more specific, IRS acts as an information carrier similar to a displayer, which is capable of showing an image like a dynamic quick response (QR) code by changing the amplitudes and phases of its elements. When a full-duplex BS transmits electromagnetic (EM) wave towards the IRS, the incident signal will be remodulated at the moment of its reflection. Once the echo signal returns to the BS, the carried data can be read. During the operation of RFCF-UT, it is clear that IRS works as a backscatter device rather than a familiar pure reflector.\par
With transmitting RF chains removed, RFCF-UT provides a transformative uplink communication way for energy- and cost-constrained terminal devices. Undoubtedly, the proposed technique has huge application potential in many scenarios. For example, low-power wireless sensors are deployed to collect and store sensing data about the time-varying surrounding environments, while a central processing equipment requests the data transfer from the sensors every once in a while~\cite{Chen2017Machine}. In this scenario, RFCF-UT may be a promising candidate technique to support long term operation of the sensor network, owing to passive uplink data migration. Another application example is vehicular networks~\cite{Liu2021Tutorial}, where a dual-functional radar-communication (DFRC) transmitter at a vehicle sends the EM wave and then receives its echo signal reflected by IRSs mounted on other ones. Through the proposed RFCF-UT technique, some information such as driving route can be acquired in addition to sensing information.  As an illustrative example for RFCF-UT, this paper will present an IRS backscatter enabled integrated sensing, communication and computation (ISCC) system. In this system, DFRC, IRS, backscatter, and data offloading are involved.
\subsection{Related Works}
Data offloading aims to migrate data bits from resource-constrained UEs to a remote resource-rich central server for computation execution. Relying on the distance from the UEs to the central server, data offloading for task computation is divided into three types: mobile cloud computing (MCC), multi-access edge computing (MEC) and multi-tier computing. For details, please refer to~\cite{Tong2016hierarchical,Haber2019Joint}. More closely related to the proposed RFCF-UT technique, we will focus on introducing DFRC, IRS and backscatter, respectively.
\subsubsection{DFRC}
Depending on whether or not radar and communication are separated, two categories of systems, respectively termed as coexisting-radar-and-communication (CRC) and DFRC, are often considered~\cite{Liu2020Joint}. For CRC systems, major researches focused on interference suppression, so as to achieve harmonious coexistence of separated radar and communication. Different from CRC, DFRC aims to simultaneously realize wireless communication and remote sensing by jointly designing a single transmitted signal. In comparison, DFRC integrates communication and sensing into one system, thereby sharing the same transmitted signal and a majority of hardware modules~\cite{Zhang2022Enabling}.
DFRC systems can be realized in multiple ways~\cite{Chen2022Generalized}. A frequently-adopted category is the multiple access technology, such as orthogonal frequency division
multiplexing waveform~\cite{Tian2017radar,Zhu2009ChunkI,Zhu2012ChunkII}, radar-aware carrier sense multiple access~\cite{Petrov2019unified}, time division multiple access~\cite{Kumari2020Adaptive}, orthogonal time frequency space~\cite{Gaudio20219Performance}, etc. Another category is to insert communication information into radar signals, such as sidelobe control~\cite{Hassanien2016Dualfunction}, code shift keying~\cite{Tedesso2018Code}, frequency-hopping code~\cite{Hassanien2019Dualfunction}, etc. Spatial beamforming design is also an efficient category to realize DFRC.
For example, Liu ~\emph{et al.}~\cite{Liu2018MUMIMO} investigated a series of transmitting beamforming methods for the DFRC systems for separated or shared antennas. Chen~\emph{et al.}~\cite{Chen2021Joint} studied beamforming design and showed a Pareto optimization framework for the considered DFRC systems. According to~\cite{Zhang2022Enabling,Liu2021Tutorial}, there are many potential application scenarios for DFRC systems, including but not limited to vehicular networks, localization, health care, and safety surveillance.
\subsubsection{IRS and Backscatter}
IRS is a cost-effective two-dimensional artificial metasurface, on which a number of passive elements are coated~\cite{Wu2020Towards}. When the radio signal impinges on IRS, each element is able to regulate the characteristics of the incident EM wave, such as amplitude and phase, in a real-time manner~\cite{Renzo2020Smart}. Therefore, the reflected signal is programmed to collaboratively achieve passive beamforming, which is utilized to enhance or weaken the reception quality at a receiver~\cite{LiuarxivMinimization}. One distinctive difference from active antennas is that IRS is not equipped with transmitting RF chains~\cite{Xu2021Envisioning,Huang2020Holographic}. Owing to the advantage of reflection passivity, many researches on IRS have been investigated, involving multi-user communications~\cite{Zhang2021MultiUser}, physical layer security (PLS)~\cite{Xu2021Intelligent}, cognitive radio~\cite{Xu2021Intelligent,Zhang2021Reconfigurable}, device-to-device (D2D) communication~\cite{Mao2021Intelligent}, multiple cells~\cite{Khan2020Centralized}, non-orthogonal multiple access~\cite{Ding2020Simple}, ect.\par
Besides the aforementioned reflection function, IRS is also able to serve as a transmitter by integrating backscatter. The key idea of backscatter techniques is to leverage existing RF signal for transmitting data instead of generating RF signals~\cite{Huynh2018Ambient}. For example, the communication signal can be embedded into the radar echo for data
transmission through backscatter~\cite{Blunt2010Intrapulse}. As a recently emerging research direction, related works on IRS backscatter or transmitter can be found in some literature, involving signal modulation~\cite{Tang2019Programmable,Tang2020MIMO,Dai2020Realization}, cognitive radio system~\cite{Guan2020Joint}, PLS~\cite{Xu2021Resisting,Xu2021IntelligentIoT}, D2D~\cite{Xu2021Envisioning}, edge computing~\cite{Xu2022Offloading}, etc.
%
%
%
%
%
%
%
\subsection{Contributions}
This paper proposes the concept of IRS backscatter enabled RFCF-UT integrally. Compared to conventional UT, the proposed RFCF-UT techniques eliminate transmitting RF chains at UEs completely, which enables substantial reduction of energy consumption and hardware complexity. Distinguished from conventional backscatter, IRS in RFCF-UT has a much larger area for EM wave reception and a higher degree of spatial freedom, thereby yielding more abundant modulation methods and a larger communication capacity. As an evolution of IRS backscatter, the ideas of IRS image modulation and coding like QR code are proposed, which need thoughtful analysis. In addition, transmitting and receiving antennas coexist at a full-duplex BS to realize a scanning function. As an illustrative example for RFCF-UT, this paper considers an IRS backscatter enabled ISCC system, with
the following contributions:
\begin{itemize}
  \item An IRS backscatter enabled ISCC system is modelled, where a DFRC BS radiates EM wave to track the radar target and provide carrier waves for the IRSs at UEs simultaneously, while the
  UEs employ the RFCF-UT technique to passively offload their raw data and computed resultant data to the BS. Furthermore, a performance metric named weighted throughput capacity (WTC) is defined to characterize the data collection capacity of system, covering sensing, communication and local computation at the UEs.
  \item This paper takes into account two data offloading strategies, namely \emph{Partial Offloading} and \emph{Binary Offloading}, according to whether the data is bitwise independent or not. Based on the strategies, two optimization problems are formulated to maximize WTC by joint optimization of the transmitting beamforming at the BS, the passive beamforming at all the IRSs, the radar receiving beamforming, the time of data computing and communication (and the integer variables). In addition to problem decomposition, linear programming (LP), fractional programming (FP), integer programming (IP) and alternative optimization (OA) are developed to solve the subproblems.
  \item The simulation results show that the proposed OA schemes for weighted sum rate for sensing and communication have a good convergence. Additionally, the achievable system WTC depends on some important parameters, including the number of elements at IRS, the transmit power at the BS, and the number of transmitting antennas at the BS, and the number of UEs. Accordingly, the feasibility and benefits of the proposed IRS backscatter enabled RFCF-UT technique are confirmed.
\end{itemize}
\subsection{Organization}
The remaining sections of this paper are presented as follows. In Section II, an IRS backscatter enabled ISCC system is modelled, followed by two optimization problem formulations. Section III and IV provide the optimization schemes, with their computational complexities. In Section V, simulation results are presented. In Section VI, the conclusions of this paper are drawn.
\section{System Model and Problem Formulation}
\begin{figure}
\centering
\includegraphics[width= 2.5in]{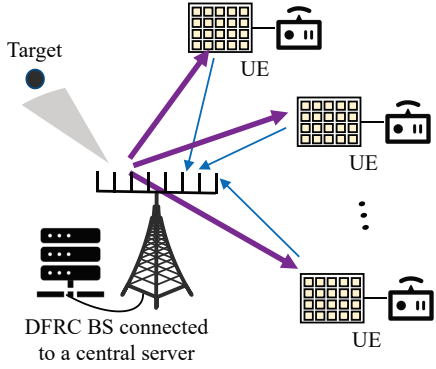}
\caption{An illustration of IRS backscatter enabled integrated sensing, communication and computation framework.}
\label{Fig2}
\end{figure}
Fig. \ref{Fig2} depicts an IRS backscatter enabled ISCC system, consisting of one $N$-antenna DFRC BS connected to a central server for data processing, $K$ identical UEs each equipped with an $L$-element IRS and a computation unit, and one target. The notations $\mathcal{K} = \{1,2,\cdots,K\}$ and  $\mathcal{L} = \{1,2,\cdots,L\}$ are used to denote the sets of all UEs and the IRS elements, respectively. By radar sensing and communication, the BS attempts to collect raw data and computed resultant data about the environment of interest. Meanwhile, the UEs are capable of harvesting the EM wave to passively offload their own data information to the BS via IRS backscatter\footnote{Since ``QR-code-like'' represents an image description of modulation for IRS backscatter or RFCF-UT, IRS backscatter is investigated to present the performance upper bound of ``QR-code-like'' modulation.}. During data offloading, RFCF uplink transmissions are executed. For the sake of illustration, only one time block of interest $T$ is considered, during which all channel states keep unchanged.\par
To realize concurrent sensing and uplink transmission, the antennas at the collocated radar and communication BS split into three groups. One group containing $N_\text{t}$ antennas acts as a DFRC transmitter, aiming to track the radar target and radiate EM wave towards the UEs simultaneously. The other two groups made up of $N_\text{r}$ and $N_\text{c}$ ($N_\text{c} > K$) antennas are used to receive the target echo signal and the information-bearing signals from the UEs, respectively. It is clear that $N = N_\text{t} + N_\text{r} + N_\text{c}$. It is assumed that there is no direct self-interference from the transmitting antennas at the BS to the receiving ones, which can be justified by physically separated antenna deployment, perfect self-interference elimination based on advanced signal processing techniques~\cite{Sabharwal2014In-band}, etc. When the EM wave transmitted by the BS impinges the IRS of each UE, new data information can be carried on it by signal modulation and then offloaded to the BS. The information consists of two types: raw data and computed resultant data, where the latter is obtained from some raw data by local computation and its size is assumed to be negligible.\par
\subsection{Communication Model}
In the network, all communication links are assumed to be flat fading. To collect the data, the BS as an energy supplier transmits the EM wave towards the UEs, and then receives the backscattered signal by their IRSs. Through this process, the data of the UEs is migrated to the BS. To be specific, let $\textbf{x} \in {\mathbb{C}^{N_\text{t} \times 1}}$ denote the transmitting beamforming vector bearing the data symbols $x$ at the BS. The dual-functional signal $\textbf{x}$ is used to realize simultaneous communication and radar sensing, with the transmit power constraint $\text{Tr}\left( \textbf{x}  \textbf{x}^H \right) \leq P$. When reaching an IRS, the signal modulation is performed to recast the signal $\textbf{x}$ for bearing new data of the associated UE. Mathematically, the signal modulation at the $k$-th IRS is given by
\begin{align}
\textbf{F}_{\text{c},k} \mathbf{\Theta}_{k}  \textbf{H}_{k} \textbf{x}
= \textbf{T}_{\text{c}, k} \boldsymbol{\theta}_{k}  \xlongequal[]{\text{modulate}}  \textbf{T}_{\text{c}, k} \textbf{v}_{k}, \nonumber
\end{align}
where $\textbf{H}_{k} \in {\mathbb{C}^{L \times N_\text{t}}}$ and $\textbf{F}_{\text{c}, k} \in {\mathbb{C}^{N_\text{c} \times L}}$ are the channel gain matrices from the transmitting antennas of the BS to the $k$-th IRS and from the $k$-th IRS to the information-receiving antennas of the BS, respectively. $\mathbf{\Theta}_{k}$ is the reflection coefficient matrix at the $k$-th IRS with $\boldsymbol{\Theta}_{k} \triangleq \text{diag} \{ \boldsymbol{\theta}_{k} \}$ and $\textbf{T}_{\text{c}, k} \triangleq \textbf{F}_{\text{c}, k} \text{diag} \{ \textbf{H}_{k} \textbf{x}\}$ defined. $\textbf{v}_{k}$ represents the information-bearing signal vector after modulation at the $k$-th IRS. From $ \boldsymbol{\theta}_{k}$ to $ \textbf{v}_{k}$, the signal modulation at the $k$-th IRS is executed to bear new data symbols $s$. By adjusting $\textbf{v}_{k}$ appropriately, the passive beamforming at the $k$-th IRS can be realized. Because the reflection coefficient at a passive IRS is no greater than one, $\left[ \textbf{v}_{k} \textbf{v}_{k}^H  \right]_{l,l}\leq 1$ holds, with $\left[\cdot \right]_{l,l}$ denoting the $l$-th diagonal element of a matrix. After modulation, the new-data-bearing signal is reflected towards the information-receiving antennas of the BS and the received signal is given by
\begin{align}
\textbf{y}_\text{c} =  \sum_{k=1}^K \textbf{T}_{\text{c}, k} \textbf{v}_{k}  + \eta \mathbf{\Lambda}_\text{c}(\theta) \textbf{x} +  \textbf{n}_\text{c}, \nonumber
\end{align}
where $\textbf{n}_\text{c}$ is a white Gaussian noise vector with $\textbf{n}_\text{c} \sim (\textbf{0}, \sigma^2 \textbf{I}_\text{c})$. $\eta$ represents the complex path-loss coefficient of the radar target located at the angle $\theta$, with $\mathbf{\Lambda}_\text{c}(\theta) \triangleq  \textbf{a}_\text{c} (\theta)  \textbf{a}_\text{t}^T(\theta)$. Clearly, $\eta \mathbf{\Lambda}_\text{c}(\theta) \textbf{x}$ represents the interference from the target echo signal. $\textbf{a}_\text{t} (\theta)$ and $\textbf{a}_\text{c}(\theta)$ are respectively given by
\begin{align}
 \textbf{a}_\text{t}(\theta) &= \frac{1}{N_\text{t}} [1, e^{j 2 \pi d \text{sin} (\theta)}, \cdots, e^{j 2 \pi (N_\text{t} - 1) d \text{sin} (\theta)}]^T, \nonumber\\
 \textbf{a}_\text{c}(\theta) &= \frac{1}{N_\text{c}} [1, e^{j 2 \pi d \text{sin} (\theta)}, \cdots, e^{j 2 \pi (N_\text{c} - 1) d \text{sin} (\theta)}]^T, \nonumber
\end{align}
where $d$ represents the spacing between two adjacent receiving antennas being normalized by the wavelength. Then, the received signal-to-interference-plus-noise ratio
(SINR) at the information-receiving antennas of the BS is given by
\begin{align}
\gamma_{\text{c},k} =  ~&\textbf{v}_{k}^H \textbf{T}_{\text{c}, k}^H \Bigg( \sigma^2 \textbf{I}_\text{c} + |\eta|^2 \mathbf{\Lambda}_\text{c}(\theta) \textbf{x} \textbf{x}^H \mathbf{\Lambda}_\text{c}^H(\theta)  \nonumber\\
& \quad\quad\quad\quad\quad\quad +  \sum_{k' \neq k}  \textbf{T}_{\text{c}, k'} \textbf{v}_{k'} \textbf{v}_{k'}^H  \textbf{T}_{\text{c}, k'}^H \Bigg) ^{-1} \textbf{T}_{\text{c}, k} \textbf{v}_{k}. \nonumber
\end{align}
%
Thus, the rate denotes given by $R_{\text{c},k} = B \log \left( 1 +   \gamma_{\text{c},k} \right)$, where $B$ is communication bandwidth. Note that when a UE does not intend to send its data to the BS, the corresponding information-bearing signal vector \textbf{v} is set as \textbf{0}. \par
\subsection{Radar Model}
Considering the radar function, the received signal at the radar-receiving antennas of the BS for given $\textbf{x}$ is expressed as
\begin{align}
\textbf{y}_\text{r} =  \eta \mathbf{\Lambda}_\text{r}(\theta) \textbf{x} + \sum_{k=1}^{K}  \textbf{T}_{\text{r}, k} \textbf{v}_{k}  + \textbf{n}_\text{r}, \nonumber
\end{align}
where $\textbf{n}_\text{r}$ is a complex Gaussian noise vector with $\textbf{n}_\text{r} \sim (\textbf{0}, \sigma^2 \textbf{I}_\text{r})$. $\mathbf{\Lambda}_\text{r}(\theta) \triangleq  \textbf{a}_\text{r} (\theta)  \textbf{a}_\text{t}^T(\theta)$ and $\textbf{T}_{\text{r}, k} \triangleq \textbf{F}_{\text{r}, k} \text{diag} \{ \textbf{H}_{k} \textbf{x}\}$. $\textbf{F}_{\text{r}, k} \in {\mathbb{C}^{N_\text{r} \times L}}$ denotes the channel gain matrix from the $k$-th IRS to the radar-receiving antennas of the BS. Clearly, $\sum_{k=1}^{K}  \textbf{T}_{\text{r}, k} \textbf{v}_{k}$ represents the interference from the reflection of all IRSs. $\textbf{a}_\text{r} (\theta)$ is given by
\begin{align}
 \textbf{a}_\text{r}(\theta) = \frac{1}{N_\text{r}} [1, e^{j 2 \pi d \text{sin} (\theta)}, \cdots, e^{j 2 \pi (N_\text{r} - 1) d \text{sin} (\theta)}]^T. \nonumber
\end{align}
Adopting the normalized receiving beamforming vector \textbf{w}, the output at the radar is given by
\begin{align}
r = \eta  \textbf{w}^H \mathbf{\Lambda}_\text{r}(\theta) \textbf{x} + \sum_{k=1}^{K}   \textbf{w}^H  \textbf{T}_{\text{r}, k} \textbf{v}_{k}  +   \textbf{w}^H  \textbf{n}_\text{r}. \nonumber
\end{align}
Accordingly, the output radar SINR is given by
\begin{align}
\gamma_\text{r} = \frac{ |\eta \textbf{w}^H \mathbf{\Lambda}_\text{r}(\theta) \textbf{x}|^2 }{\sum_{k=1}^{K}   |\textbf{w}^H  \textbf{T}_{\text{r}, k} \textbf{v}_{k}|^2  +  \sigma^2  |\textbf{w}|^2}, \nonumber
\end{align}
\subsection{Computation and Energy Model}
In this network, each UE with separated computing and communicating circuit units requests to send its data to the BS. The data is divided into two types: raw data and computed resultant data. Generally speaking, the latter has an extremely small size owing to local computation, hence its migration time from a UE to the BS is reasonably ignored. Let $f_k$, $c_k$ and $\varepsilon_k$ denote the CPU's frequency, the cycle number and the energy consumption coefficient of processor's chip for computing one bit at each UE, respectively. The computational rate at each UE is given by
\begin{align}
R_{\text{loc}, k} = \frac{f_k}{c_k}. \nonumber
\end{align}
The energy consumption at the $k$-th UE is made up of computing data bits and running IRS, which is given by
\begin{align}
E_{k} = t_{\text{loc}, k} \varepsilon_k f_k^3 + t_{\text{c}} L \mu, \nonumber
\end{align}
where $t_{\text{loc}, k}$ is the data computing time locally at the $k$-th UE. $t_{\text{c}}$ represents the communication time, during which the IRSs and the BS are all in on-state. $\mu$ denotes each IRS element's power consumption, positively associated with its phase resolution. For an IRS, an increase in the element number must result in more power consumption. On the other hand, the central server is assumed to have an infinite capacity of data computing and thus the computation time for the data from the UEs is ingored. In this paper, we consider two data offloading strategies. 1) \emph{Partial Offloading}: Assuming that the data is bitwise independent, all bits split into two subsets. One as raw data is directly sent to the BS, while the other is locally computed and then its resultant data is sent to the BS. 2) \emph{Binary Offloading}: All the data bits are either directly offloaded to the BS, or locally transformed into the resultant data that is subsequently migrated to the BS.
\subsection{Problem Formulation}
In the considered model, the BS attempts to extract as much information as possible from the environment of interest. To evaluate the data collection capacity of system, a performance metric named WTC is defined as follows.
\begin{align}
C_{\text{po}} &=  \omega_\text{r} t_\text{c} R_\text{r} + \sum_{k=1}^K  \omega_{k} \left( t_{\text{c}} R_{\text{c},k} +  t_{\text{loc}, k} R_{\text{loc}, k} \right), \nonumber\\
C_{\text{bo}} &= \omega_\text{r} t_\text{c} R_\text{r} + \sum_{k=1}^K  \omega_{k} \left[ \xi_k t_{\text{c}} R_{\text{c},k} + (1-\xi_k) t_{\text{loc}, k} R_{\text{loc}, k} \right], \nonumber
\end{align}
where $C_{\text{po}}$ and $C_{\text{bo}}$ are the WTC for \emph{Partial Offloading} and \emph{Binary Offloading}, respectively. For $R_{\text{r}} = B \log \left( 1 +   \gamma_{\text{r}} \right) $, it is used to characterize the radar sensing capacity. $\xi_k$ are binary integers, with $\xi_k \in \{0,1\}$. In this paper, we take into account the maximization problem of WTC by joint optimization of the transmitting beamforming at the BS, the passive beamforming at all the IRSs, the radar receiving beamforming, the time of data computing and communication (and the integer variables). \par
1) For the \emph{Partial Offloading} strategy, the optimization problem is formulated as
\begin{subequations}
\begin{align}
(\text{P1})\quad \underset{\mathcal{V}, \mathcal{T}, \textbf{x}, \textbf{w}} \max \quad &  C_{\text{po}}, \nonumber \\
s.t. \quad
 & \text{C1}: \text{Tr}\left( \textbf{x} \textbf{x}^H \right) \leq P,  \nonumber\\
 & \text{C2}: \left[ \textbf{v}_{k} \textbf{v}_{k}^H \right]_{l,l}\leq 1, ~k \in \mathcal{K}, l \in \mathcal{L}, \nonumber\\
 & \text{C3}: \text{Tr}\left( \textbf{w} \textbf{w}^H \right) \leq 1, \nonumber\\
 & \text{C4}: E_{k} \leq E_{k}^{\text{th}}, ~k \in \mathcal{K}, \nonumber\\
 & \text{C5}: t_{\text{loc}, k} \leq T,   ~k \in \mathcal{K},  \nonumber\\
 & \text{C6}: t_{\text{c}} \leq T, \nonumber
\end{align}
\end{subequations}
\par 2) For the \emph{Binary Offloading} strategy, the optimization problem is formulated as
\begin{subequations}
\begin{align}
(\text{P2})\quad \underset{\mathcal{V}, \mathcal{T}, \textbf{x}, \textbf{w}, \Xi} \max \quad &  C_{\text{bo}}, \nonumber \\
s.t. \quad
 & \text{C1} \sim \text{C6}, \nonumber\\
 & \text{C7}: \frac{K}{2} \leq \sum_{k=1}^K \xi_k \leq \frac{K+1}{2}, \nonumber\\
 & \text{C8}: \xi_k \in \{0,1\}, ~k \in \mathcal{K}. \nonumber
\end{align}
\end{subequations}
\noindent where $\mathcal{V}$ and $\Xi$ refer to the collections of $\textbf{v}_k$ and $\xi_k$, respectively. $\mathcal{T}$ denotes the collection of $t_{\text{loc}, k}$ and $t_{\text{c}}$.  C1 represents the transmitting power budget constraint at the BS; C2 denotes the complex reflection coefficient constraint at all the IRSs, involving amplitude and phase shift of each element; C3 is the constraint of the receiving beamforming vector of radar antennas; C4 denotes the energy constraint for each IRS backscatter assisted UE, where $E_{k}^{\text{th}}$ is the energy threshold value of the $k$-th UE with $E_{k}^{\text{th}} > T L \mu$ generally considered; C5 and C6 denote the time constraints; C7 and C8 denote the user scheduling constraints for uplink transmission from the UEs to the BS, or local computation. Through the constraints C7 and C8, it is set that approximately half of UEs can offload their data to the BS in one time slot.
\section{WTC Maximization for Partial Offloading}
This section focuses on the optimization of the problem (P1) for the \emph{Binary Offloading} strategy. In this strategy, the problem (P1) is rewritten as
\begin{subequations}
\begin{align}
\underset{\mathcal{V}, \mathcal{T}, \textbf{x}, \textbf{w}} \max \quad & \omega_\text{r} t_\text{c} R_\text{r} + \sum_{k=1}^K  \omega_{k} \left( t_{\text{c}} R_{\text{c},k} +  t_{\text{loc}, k} R_{\text{loc}, k} \right), \nonumber \\
s.t. \quad
 & \text{C1} \sim \text{C6}. \nonumber
\end{align}
\end{subequations}
Due to deep coupling of the optimization variables $\mathcal{V}$, $\mathcal{T}$, $\textbf{x}$ and $\textbf{w}$, it is quite challenging to address this problem directly. Moreover, logarithm functions are involved in the objective function, which further raises the difficulty level in solving this problem. Fortunately, this problem for given $\omega_\text{r} R_\text{r}$ and $\sum_{k=1}^K  \omega_{k} R_{\text{c}, k}$ is in a form of linear programming, which depends only on $\mathcal{T}$ related to the constraints C4, C5 and C6. On the other hand, $R_{\text{c}, k}$ and $R_\text{r}$ are functions of the optimization variables $\mathcal{V}$, $\textbf{x}$ and $\textbf{w}$, which are only involved in the constraints C1, C2 and C3. Therefore, this problem is decomposed into two subproblems (P1.1) and (P1.2), which are respectively given by
\begin{subequations}
\begin{align}
(\text{P1.1})\quad \underset{\mathcal{T}} \max \quad & t_\text{c} \left( \omega_\text{r} R_\text{r} + \sum_{k=1}^K  \omega_{k} R_{\text{c},k} \right) + \sum_{k=1}^K  t_{\text{loc}, k} \omega_{k} R_{\text{loc}, k}, \nonumber \\
s.t. \quad
 & \text{C4}, \text{C5}, \text{C6}, \nonumber
\end{align}
\end{subequations}
and
\begin{subequations}
\begin{align}
(\text{P1.2})\quad \underset{\mathcal{V}, \textbf{x}, \textbf{w}} \max \quad &  \omega_\text{r} R_\text{r} + \sum_{k=1}^K \omega_{k} R_{\text{c},k}, \nonumber \\
s.t. \quad
 & \text{C1}, \text{C2}, \text{C3}.  \nonumber
\end{align}
\end{subequations}
Since the linear programming problem (P1.1) for given $\omega_\text{r} R_\text{r}$ and $\sum_{k=1}^K  \omega_{k} R_{\text{c}, k}$ is easy to address, we will focus only on the problem (P1.2). To make the problem (P1.2) feasible, Lagrangian dual transform as a frequently-used fractional programming (FP) method is adopted to remould its objective function. Based on this, the optimization variables $\mathcal{V}$, $\textbf{x}$ and $\textbf{w}$ are alternatively optimized.
\subsection{Remoulding Objective Function}
From the problem (P1.2), it is easily seen that its objective function $\omega_\text{r} R_\text{r} + \sum_{k=1}^K \omega_{k} R_{\text{c},k}$ is a weighted sum of $K+1$ logarithm functions, resulting in more difficulty in finding the optimal solution. To deal with such a difficulty, the objective function is firstly changed into a more tractable form by Lagrangian dual transform. To be specific, the weighted sum of logarithm functions $\omega_\text{r} R_\text{r} + \sum_{k=1}^K \omega_{k} R_{\text{c},k}$ can be rewritten as
\begin{align}
& \omega_\text{r} R_\text{r} + \sum_{k=1}^K \omega_{k} R_{\text{c},k}
=  \omega_\text{r} \log \left( 1 +   \gamma_{\text{r}} \right) + \sum_{k=1}^K   \omega_{k} \log \left( 1 +   \gamma_{\text{c},k} \right)  \nonumber \\
 &=\underset{\alpha_\text{r}, \alpha_{k} \geq 0} \max ~   \omega_{r} \left[ \log \left( 1 + \alpha_\text{r} \right) - \alpha_\text{r} \right]  +  \frac{\omega_\text{r}  \left( 1 + \alpha_\text{r} \right) \gamma_{\text{r}} }{ 1 + \gamma_{\text{r}}} \nonumber\\
 & \quad\quad\quad + \sum_{k=1}^K  \omega_{k} \left[ \log \left( 1 + \alpha_{k} \right) - \alpha_{k} \right]  +  \frac{\omega_{k}  \left( 1 + \alpha_{k} \right) \gamma_{\text{c},k} }{ 1 + \gamma_{\text{c},k}}, \nonumber
\end{align}
where $\alpha_\text{r}$ and  $\alpha_{k}$ are auxiliary variables. According to the quadratic transform given in ~\cite{Shen2018Fractional}, it is derived that
\begin{align}
& \frac{\omega_\text{r}  \left( 1 + \alpha_\text{r} \right) \gamma_{\text{r}} }{ 1 + \gamma_{\text{r} }} = \frac{\omega_\text{r}  \left( 1 + \alpha_\text{r} \right)  | \mathrm{A}_{\text{r}}|^2 }{\mathrm{B}_\text{r}}  \nonumber\\
= & 2 \sqrt{ \omega_\text{r}( 1 + \alpha_\text{r}) }  \text{Re} \{{\beta}_\text{r}^\ast \mathrm{A}_{\text{r}} \}  -    {\beta}_\text{r}^\ast \mathrm{B}_\text{r}  {\beta}_\text{r}, \nonumber \\
& \frac{\omega_{k}  \left( 1 + \alpha_{k} \right) \gamma_{\text{c},k} }{ 1 + \gamma_{\text{c},k}} = \frac{\omega_{k}  \left( 1 + \alpha_{k} \right)  | \mathrm{A}_{\text{c},k}|^2 }{\mathrm{B}_\text{c}}  \nonumber\\
= & 2 \sqrt{ \omega_{k}( 1 + \alpha_{k}) }  \text{Re} \{ \boldsymbol{\beta}_{k}^H \mathrm{A}_{\text{c},k} \}  -    \boldsymbol{\beta}_{k}^H \mathrm{B}_\text{c}  \boldsymbol{\beta}_{k}, \nonumber
\end{align}
where ${\beta}_\text{r}$ and $\boldsymbol{\beta}_{k}$ represent auxiliary variable scalar or vectors. $\mathrm{A}_\text{r}$, $\mathrm{B}_\text{r}$, $\mathrm{A}_{\text{c},k}$ and $\mathrm{B}_\text{c}$ are respectively given by
\begin{align}
& \mathrm{A}_\text{r} =  \eta  \textbf{w}^H \mathbf{\Lambda}_\text{r}(\theta) \textbf{x}, \nonumber\\
& \mathrm{B}_\text{r} = \sigma^2 |\textbf{w}|^2 +  |\eta \textbf{w}^H \mathbf{\Lambda}_\text{r}(\theta) \textbf{x}|^2 + \sum_{k'=1}^{K}   |\textbf{w}^H  \textbf{T}_{\text{r}, k'} \textbf{v}_{k'}|^2,  \nonumber\\
& \mathrm{A}_{\text{c},k} =  \textbf{T}_{\text{c}, k} \textbf{v}_{k}, \nonumber\\
& \mathrm{B}_\text{c} = \sigma^2 \textbf{I}_\text{c} + |\eta|^2 \mathbf{\Lambda}_\text{c}(\theta) \textbf{x} \textbf{x}^H \mathbf{\Lambda}_\text{c}^H(\theta) +  \sum_{k'=1}^K  \textbf{T}_{\text{c}, k'} \textbf{v}_{k'} \textbf{v}_{k'}^H  \textbf{T}_{\text{c}, k'}^H.  \nonumber
\end{align}
Thus, the problem (\text{P1.2}) is rewritten as
\begin{align}
(\text{P1.3}) \quad \underset{\mathcal{V}, \textbf{x}, \textbf{w}, \alpha, \boldsymbol{\beta}} \max \quad &   f(\mathcal{V}, \textbf{x}, \textbf{w}, \alpha, \boldsymbol{\beta}), \nonumber\\
 & \text{C1}, \text{C2}, \text{C3}, \nonumber\\
 & \text{C4}: \alpha_\text{r}\geq 0, \alpha_{k} \geq 0, k \in \mathcal{K}, \nonumber
\end{align}
where $\alpha$ represents the collection of the auxiliary variables $\alpha_\text{r}$ and $\alpha_{k}$. $\boldsymbol{\beta}$ is the collection of the auxiliary variable scalar ${\beta}_\text{r}$ and vectors $\boldsymbol{\beta}_{k}$. $f(\mathcal{V}, \textbf{x}, \textbf{w}, \alpha, \boldsymbol{\beta})$ is given by
\begin{align}
f(\mathcal{V}, \textbf{x}, \textbf{w}, \alpha, \boldsymbol{\beta}) = & \omega_{r} \left[ \log \left( 1 + \alpha_\text{r} \right) - \alpha_\text{r} \right] \nonumber\\
& + 2 \sqrt{ \omega_\text{r}( 1 + \alpha_\text{r}) }  \text{Re} \{{\beta}_\text{r}^\ast \mathrm{A}_{\text{r}} \}  -    {\beta}_\text{r}^\ast \mathrm{B}_\text{r}  {\beta}_\text{r} \nonumber\\
& + \sum_{k=1}^K  \omega_{k} \left[ \log \left( 1 + \alpha_{k} \right) - \alpha_{k} \right]  \nonumber\\
& + 2 \sqrt{ \omega_{k}( 1 + \alpha_{k}) }  \text{Re} \{ \boldsymbol{\beta}_{k}^H \mathrm{A}_{\text{c},k} \}  -    \boldsymbol{\beta}_{k}^H \mathrm{B}_\text{c}  \boldsymbol{\beta}_{k}. \nonumber
\end{align}
In the problem (P1.3), the optimization variables $\mathcal{V}$, $\textbf{x}$, $\textbf{w}$, $\alpha$ and $\boldsymbol{\beta}$ are deeply coupled in the objective function and the constraints. In response to such a challenge, we will employ the AO method to iteratively seek the optimal optimization variables.
\subsection{Alternative Optimization}
To address the problem (P1.3), an AO procedure is provided to cyclically optimize the variables $\mathcal{V}$, $\textbf{x}$, $\textbf{w}$, $\alpha$ and $\boldsymbol{\beta}$, which is decomposed into four steps.\par
\vspace{0.1cm}
\noindent \emph{Step-1): Optimizing} $\alpha$ \emph{and} $\boldsymbol{\beta}$ \par
\vspace{0.1cm}
With $\mathcal{V}$, $\textbf{x}$ and $\textbf{w}$ given, we take a derivative with respect to $\alpha_\text{r}$, $\alpha_k$, $\beta_\text{r}$ and $\boldsymbol{\beta}_k$ to obtain the optimal $\alpha_\text{r}^\circ$, $\alpha_k^\circ$, $\beta_\text{r}^\circ$ and $\boldsymbol{\beta}_k^\circ$, separately. Mathematically, let
\setcounter{equation}{0}
\begin{align}
& \frac{\partial   f(\mathcal{V}, \textbf{x}, \textbf{w}, \alpha, \boldsymbol{\beta}) }{\partial \alpha_\text{r}} = 0,   \label{Eq1} \\
& \frac{\partial   f(\mathcal{V}, \textbf{x}, \textbf{w}, \alpha, \boldsymbol{\beta}) }{\partial {\beta}_\text{r}} = 0,   \label{Eq2}\\
& \frac{\partial   f(\mathcal{V}, \textbf{x}, \textbf{w}, \alpha, \boldsymbol{\beta}) }{\partial \alpha_{k}} = 0,    \label{Eq3}\\
& \frac{\partial   f(\mathcal{V}, \textbf{x}, \textbf{w}, \alpha, \boldsymbol{\beta}) }{\partial \boldsymbol{\beta}_{k}} = 0.  \label{Eq4}
\end{align}
It is not difficult to find out the optimal $\alpha_\text{r}^\circ$, $\alpha_k^\circ$, $\beta_\text{r}^\circ$ and $\boldsymbol{\beta}_k^\circ$, which are respectively given by
\begin{align}
\alpha_\text{r}^\circ &=  \gamma_{\text{r}} , \nonumber\\
{\beta}_\text{r}^\circ &=  \sqrt{ \omega_\text{r}( 1 + \alpha_\text{r}) } {\mathrm{B}_\text{r}^{-1}}{ \mathrm{A}_{\text{r}}}, \nonumber\\
\alpha_{k}^\circ &=  \gamma_{\text{c},k} , \nonumber\\
\boldsymbol{\beta}_{k}^\circ &=  \sqrt{ \omega_{k}( 1 + \alpha_{k}) } {\mathrm{B}_\text{c}^{-1}}{ \mathrm{A}_{\text{c},k}}. \nonumber
\end{align}
\vspace{0.1cm}
\noindent \emph{Step-2):} Optimizing $\textbf{x}$ \par
\vspace{0.1cm}
Given $\alpha$, $\boldsymbol{\beta}$, $\mathcal{V}$, and $\textbf{w}$ , the objective function maximization of the problem (\text{P1.3}) is equivalent to
\begin{subequations}
\begin{align}
&\underset{\textbf{x}} \max \quad   f(\mathcal{V}, \textbf{x}, \textbf{w}, \alpha, \boldsymbol{\beta})  \nonumber \\
\Longleftrightarrow & \underset{\textbf{x}} \min \quad  \textbf{x}^H  \textbf{Y} \textbf{x} - 2 \text{Re} \{ \textbf{x}^H \textbf{z} \}, \nonumber
\end{align}
\end{subequations}
where
\begin{align}
& \textbf{Y} = \textbf{Y}_\text{r} + \textbf{Y}_\text{c}, \quad \textbf{z} = \textbf{z}_\text{r} + \textbf{z}_\text{c}, \nonumber\\
& \textbf{Y}_\text{r} =  |\eta|^2 {\beta}_\text{r}^\ast{\beta}_\text{r} \mathbf{\Lambda}_\text{r}^H(\theta) \textbf{w} \textbf{w}^H \mathbf{\Lambda}_\text{r}(\theta) + \nonumber\\
& \quad\quad \sum_{k'=1}^K   ({\beta}_\text{r}^\ast \textbf{w}^H \textbf{F}_{\text{r}, k'}   \text{diag} \{ \textbf{v}_{k'} \}  \textbf{H}_{k'} )^H {\beta}_\text{r}^\ast \textbf{w}^H \textbf{F}_{\text{r}, k'}   \text{diag} \{ \textbf{v}_{k'} \} \textbf{H}_{k'},  \nonumber\\
& \textbf{z}_\text{r} =  \sqrt{ \omega_\text{r} ( 1 + \alpha_\text{r}) }  ({\beta}_\text{r}^\ast \eta  \textbf{w}^H \mathbf{\Lambda}_\text{r}(\theta) )^H,  \nonumber\\
& \textbf{Y}_\text{c} = \sum_{k=1}^K \Big [ |\eta|^2 \mathbf{\Lambda}_\text{c}^H(\theta) \boldsymbol{\beta}_{k}  \boldsymbol{\beta}_{k}^H \mathbf{\Lambda}_\text{c}(\theta) \nonumber\\
& \quad\quad + \sum_{k'=1}^K   (\boldsymbol{\beta}_{k}^H \textbf{F}_{\text{c}, k'}   \text{diag} \{ \textbf{v}_{k'} \}  \textbf{H}_{k'} )^H  \boldsymbol{\beta}_{k}^H \textbf{F}_{\text{c}, k'}   \text{diag} \{ \textbf{v}_{k'} \} \textbf{H}_{k'} \Big],  \nonumber\\
& \textbf{z}_\text{c} = \sum_{k=1}^K  \sqrt{ \omega_{k}( 1 + \alpha_{k}) }  (\boldsymbol{\beta}_{k}^H \textbf{F}_{\text{c}, k}  \text{diag} \{ \textbf{v}_k \} \textbf{H}_{k} )^H,  \nonumber
\end{align}
Therefore, the problem (\text{P1.3}) is reformulated as
\begin{subequations}
\begin{align}
& \underset{\textbf{x}} \min \quad  \textbf{x}^H  \textbf{Y} \textbf{x} - 2 \text{Re} \{ \textbf{x}^H \textbf{z} \}, \nonumber\\
& s.t. \quad
 \text{Tr}(\textbf{x} \textbf{x}^H) \leq P. \nonumber
\end{align}
\end{subequations}
The associated Lagrangian is given by
\begin{align}
\mathcal{L}_1(\textbf{x},\lambda_1) = \textbf{x}^H  \textbf{Y} \textbf{x} - 2 \text{Re} \{ \textbf{x}^H \textbf{z} \}  +  \lambda_1 \left( \textbf{x}^H \textbf{I} \textbf{x} - P \right), \nonumber
\end{align}
where $\lambda_1$ denotes the Lagrange multiplier. By setting $ \frac{\partial \mathcal{L}_1(\textbf{x},\lambda_1)}{\partial \textbf{x}} =0 $, the optimal $\textbf{x}^\circ$ is given by
\setcounter{equation}{4}
\begin{align}
& \textbf{x}^\circ =  \left(  \lambda_1 \textbf{I}  +  \textbf{Y} \right)^{-1} \textbf{z},   \label{Eq5}\\
& \lambda_1 ^\circ =  \min \left\{ \lambda_1 \geq 0:   \textbf{x}^H \textbf{I} \textbf{x} \leq P \right\}.  \label{Eq6}
\end{align}
\noindent \emph{Step-3): Optimizing} $\mathcal{V}$ \par
\vspace{0.1cm}
\par Given $\alpha$, $\boldsymbol{\beta}$, $\textbf{x}$, and $\textbf{w}$, the objective function maximization of the problem (\text{P1.3}) is equivalent to
\begin{align}
&\underset{\mathcal{V}} \max \quad   f(\mathcal{V}, \textbf{x}, \textbf{w}, \alpha, \boldsymbol{\beta})  \nonumber \\
\Longleftrightarrow & \underset{\mathcal{V}} \max  \quad     2 \sqrt{ \omega_\text{r}( 1 + \alpha_\text{r}) }  \text{Re} \{{\beta}_\text{r}^\ast \mathrm{A}_{\text{r}} \}  -    {\beta}_\text{r}^\ast \mathrm{B}_\text{r}  {\beta}_\text{r} \nonumber\\
& \quad\quad\quad + \sum_{k=1}^K  2 \sqrt{ \omega_{k}( 1 + \alpha_{k}) }  \text{Re} \{ \boldsymbol{\beta}_{k}^H \mathrm{A}_{\text{c},k} \}  -    \boldsymbol{\beta}_{k}^H \mathrm{B}_\text{c}  \boldsymbol{\beta}_{k}. \nonumber\\
\Longleftrightarrow &  \underset{\mathcal{V}} \max  \quad   -  |{\beta}_\text{r}|^2 \left( \sum_{k'=1}^K \textbf{w}^H  \textbf{T}_{\text{r}, k'} \textbf{v}_{k'} \textbf{v}_{k'}^H \textbf{T}_{\text{r}, k'} ^H \textbf{w} \right), \nonumber\\
 & \quad\quad\quad \sum_{k=1}^K 2 \sqrt{ \omega_{k}( 1 + \alpha_{k}) }  \text{Re} \{ \boldsymbol{\beta}_{k}^H \textbf{T}_{\text{c}, k} \textbf{v}_{k} \}  \nonumber\\
 & \quad\quad\quad\quad\quad\quad\quad\quad  -    \boldsymbol{\beta}_{k}^H\left( \sum_{k'=1}^K \textbf{T}_{\text{c}, k'} \textbf{v}_{k'} \textbf{v}_{k'}^H \textbf{T}_{\text{c}, k'}^H \right) \boldsymbol{\beta}_{k}. \nonumber
\end{align}
It is not difficult to derive that
\begin{align}
& |{\beta}_\text{r}|^2 \left( \sum_{k'=1}^K \textbf{w}^H  \textbf{T}_{\text{r}, k'} \textbf{v}_{k'} \textbf{v}_{k'}^H \textbf{T}_{\text{r}, k'} ^H \textbf{w} \right) =  \sum_{k'=1}^K \text{Tr} \left(|{\beta}_\text{r}|^2 \textbf{T}_{\text{r}, k'} ^H \textbf{W}  \textbf{T}_{\text{r}, k'} \textbf{V}_{k'}    \right), \nonumber\\
&2  \text{Re} \{  \boldsymbol{\beta}_{k}^H \textbf{T}_{\text{c}, k} \textbf{v}_{k} \} = \text{Tr} (\mathbf{\Omega}_{k}   \hat{\textbf{V}}_{k}),  \nonumber\\
&   \boldsymbol{\beta}_{k}^H\left( \sum_{k'=1}^K \textbf{T}_{\text{c}, k'} \textbf{v}_{k'} \textbf{v}_{k'}^H \textbf{T}_{\text{c}, k'}^H \right) \boldsymbol{\beta}_{k} =  \sum_{k'=1}^K \text{Tr} \left( \boldsymbol{\beta}_{k} \boldsymbol{\beta}_{k}^H {\textbf{T}}_{\text{c}, k'} \textbf{V}_{k'} {\textbf{T}}_{\text{c}, k'}^H \right),  \nonumber
\end{align}
where $\textbf{V}_{k} \triangleq \textbf{v}_{k} \textbf{v}_{k}^H $, $\hat{\textbf{V}}_{k} \triangleq \hat{\textbf{v}}_{k} \hat{\textbf{v}}_{k}^H $, and
\begin{align} \nonumber
\mathbf{\Omega}_{k} =    \left[ \begin{matrix}
   \textbf{0}  &\left( \boldsymbol{\beta}_{k}^H \textbf{T}_{\text{c}, k} \right)^H\\
   \boldsymbol{\beta}_{k}^H \textbf{T}_{\text{c}, k}   &0
  \end{matrix}\right], ~
\hat{\textbf{v}}_{k} =  \left[ \begin{matrix}
   \textbf{v}_{k}\\
   1
  \end{matrix}\right]. \nonumber
\end{align}
Based on this, the new objective function is given by
\begin{align} \nonumber
&\underset{\hat{\textbf{V}}_{k}} \max \quad   f(\mathcal{V}, \textbf{x}, \textbf{w}, \alpha, \boldsymbol{\beta})  \nonumber \\
\Longleftrightarrow &
\underset{\hat{\textbf{V}}_{k}} \max  \quad  - \sum_{k'=1}^K \text{Tr} \left(|{\beta}_\text{r}|^2 \textbf{T}_{\text{r}, k'} ^H \textbf{W}  \textbf{T}_{\text{r}, k'} \textbf{V}_{k'}    \right)  \nonumber\\
& \quad\quad \quad \sum_{k=1}^K  \Bigg[ \sqrt{ \omega_{k}( 1 + \alpha_{k}) }   \text{Tr} (\mathbf{\Omega}_{k}   \hat{\textbf{V}}_{k})  \nonumber\\
 &  \quad\quad\quad -    \sum_{k'=1}^K \text{Tr} \left( \boldsymbol{\beta}_{k} \boldsymbol{\beta}_{k}^H {\textbf{T}}_{\text{c}, k'} \textbf{V}_{k'} {\textbf{T}}_{\text{c}, k'}^H \right) \Bigg]. \nonumber
\end{align}
Thus, the problem (P1.3) is rewritten as
\begin{align}
(\text{P1.4}) \quad \underset{\hat{\textbf{V}}_{k}} \max \quad &- \sum_{k'=1}^K \text{Tr} \left(|{\beta}_\text{r}|^2 \textbf{T}_{\text{r}, k'} ^H \textbf{W}  \textbf{T}_{\text{r}, k'} \textbf{V}_{k'}    \right)  \nonumber\\
&   + \sum_{k=1}^K  \Bigg[ \sqrt{ \omega_{k}( 1 + \alpha_{k}) }   \text{Tr} (\mathbf{\Omega}_{k}   \hat{\textbf{V}}_{k})  \nonumber\\
 &   -    \sum_{k'=1}^K \text{Tr} \left( \boldsymbol{\beta}_{k} \boldsymbol{\beta}_{k}^H {\textbf{T}}_{\text{c}, k'} \textbf{V}_{k'} {\textbf{T}}_{\text{c}, k'}^H \right) \Bigg], \nonumber\\
s.t. \quad
 & \text{C9}: \left[ \hat{\textbf{V}}_{k} \right]_{l,l}\leq 1, ~l \in \mathcal{L}, k \in \mathcal{K}, \nonumber\\
 & \text{C10}: \left[  \hat{\textbf{V}}_{k} \right]_{L+1,L+1} = 1,  k \in \mathcal{K}, \nonumber\\
 & \text{C11}:  \hat{\textbf{V}}_{k} \succeq \textbf{0}, k \in \mathcal{K}, \nonumber\\
 & \text{C12}: \text{rank}\left(  \hat{\textbf{V}}_{k} \right) = 1, k \in \mathcal{K}, \nonumber
\end{align}
Ignoring the constraint C12, the problem (P1.4) is relaxed as
\begin{align}
(\text{P1.5}) \quad \underset{\hat{\textbf{V}}_{k}} \max \quad &- \sum_{k'=1}^K \text{Tr} \left(|{\beta}_\text{r}|^2 \textbf{T}_{\text{r}, k'} ^H \textbf{W}  \textbf{T}_{\text{r}, k'} \textbf{V}_{k'}    \right)  \nonumber\\
&   + \sum_{k=1}^K  \Bigg[ \sqrt{ \omega_{k}( 1 + \alpha_{k}) }   \text{Tr} (\mathbf{\Omega}_{k}   \hat{\textbf{V}}_{k})  \nonumber\\
 &  -    \sum_{k'=1}^K \text{Tr} \left( \boldsymbol{\beta}_{k} \boldsymbol{\beta}_{k}^H {\textbf{T}}_{\text{c}, k'} \textbf{V}_{k'} {\textbf{T}}_{\text{c}, k'}^H \right) \Bigg], \nonumber\\
s.t. \quad
 & \text{C9} \sim \text{C11}. \nonumber
\end{align}
Clearly, (\text{P1.5}) is a convex semidefinite programming (SDP) problem over $\hat{\textbf{V}}_{k}$ and easy to solve. Then, the corresponding rank-one solution $\textbf{v}_{k}$ can be recovered by using singular value decomposition (SVD), the eigenvector corresponding to the maximum eigenvalue, or the Gaussian randomization method. \par
\noindent \emph{Step-4): Optimizing} $\textbf{w}$ \par
\par Given $\alpha$, $\boldsymbol{\beta}$, $\textbf{x}$, and $\mathcal{V}$,  the objective function maximization of the problem (\text{P1.3}) is equivalent to
\begin{align}
\underset{\textbf{w}} \max \quad   f(\mathcal{V}, \textbf{x}, \textbf{w}, \alpha, \boldsymbol{\beta})
\Longleftrightarrow   \underset{\textbf{w}} \min  \quad  \textbf{w}^H \textbf{Q} \textbf{w} - 2 \text{Re} \{ \textbf{w}^H \textbf{p} \}, \nonumber
\end{align}
where
\begin{align}
\textbf{p} &=  \sqrt{ \omega_\text{r}( 1 + \alpha_\text{r}) } {\beta}_\text{r}^\ast \eta  \mathbf{\Lambda}_\text{r}(\theta) \textbf{x}, \nonumber \\
\textbf{Q} &=  |{\beta}_\text{r}|^2 \left( |\eta|^2 \mathbf{\Lambda}_\text{r}(\theta) \textbf{x} \textbf{x}^H \mathbf{\Lambda}_\text{r}(\theta)^H + \sum_{k'=1}^{K}   \textbf{T}_{\text{r}, k'} \textbf{v}_{k'} \textbf{v}_{k'}^H \textbf{T}_{\text{r}, k'}^H \right). \nonumber
\end{align}
Therefore, the problem (\text{P1.3}) is reformulated as
\begin{subequations}
\begin{align}
& \underset{\textbf{w}} \min \quad \textbf{w}^H \textbf{Q} \textbf{w} - 2 \text{Re} \{ \textbf{w}^H \textbf{p} \}, \nonumber\\
& s.t. \quad
 \text{Tr}(\textbf{w} \textbf{w}^H) \leq 1. \nonumber
\end{align}
\end{subequations}
The associated Lagrangian is given by
\begin{align}
\mathcal{L}_2(\textbf{w},\lambda_2) = \textbf{w}^H \textbf{Q} \textbf{w} - 2 \text{Re} \{ \textbf{w}^H \textbf{p}\} +  \lambda_2 \left( \textbf{w}^H \textbf{I} \textbf{w} - 1 \right), \nonumber
\end{align}
where $\lambda_2$ denotes the Lagrange multiplier. By setting $ \frac{\partial \mathcal{L}_2 (\textbf{w},\lambda_2)}{\partial \textbf{w}} =0 $, the optimal $\textbf{w}^\circ$ is given by
\setcounter{equation}{6}
\begin{align}
& \textbf{w}^\circ =  \left(  \lambda_2 \textbf{I}  +  \textbf{Q} \right)^{-1} \textbf{v},   \label{Eq7} \\
& \lambda_2^\circ =  \min \left\{ \lambda_2 \geq 0:   \textbf{w}^H \textbf{I} \textbf{w} \leq 1 \right\}.   \label{Eq8}
\end{align}
\subsection{Complexity Analysis}
\begin{algorithm}[t]
\caption{Overall algorithm for the problem (P1)}
\begin{algorithmic}[1]
\State \textbf{Initialization}: Set $m$ = 0, $\varepsilon$, $\textbf{x}$, $\textbf{w}$, $\mathcal{V}$, $f_\text{opt}^{(m)}$.
\State \textbf{repeat}
\State \quad Set $m = m + 1$.
\State \quad Compute $\alpha_\text{r}^{(m)}$, $\alpha_k^{(m)}$, $\beta_\text{r}^{(m)}$ and $\boldsymbol{\beta}_k^{(m)}$ by \eqref{Eq1} $\sim$ \eqref{Eq4}. \nonumber
\State \quad  Find $\textbf{x}^{(m)}$ by \eqref{Eq5} and \eqref{Eq6}. \nonumber
\State \quad  Solve (\text{P1.5}) and employ rank-one recovery to obtain $\mathcal{V}^{(m)}$. \nonumber
\State \quad  Find $\textbf{w}^{(m)}$ by \eqref{Eq7} and \eqref{Eq8}. \nonumber
\State \quad  Compute $f_\text{opt}^{(m)} = f(\mathcal{V}^{(m)}, \textbf{x}^{(m)}, \textbf{w}^{(m)}, \alpha^{(m)}, \boldsymbol{\beta}^{(m)}) $. \nonumber
\State \textbf{until} $\frac{f_\text{opt}^{(m)} - f_\text{opt}^{(m-1)}}{f_\text{opt}^{(m)}} < \varepsilon$.
\State Employ LP to solve (\text{P1.1}) to obtain the optimal $C_{\text{po}}^{\ast}$.
\State \textbf{return} $C_{\text{po}}^{\ast}$.
\end{algorithmic}
\end{algorithm}
The overall algorithm for the problem (P1) is given in Algorithm 1. The problem (P1) can be solved by its two subproblems (P1.1) and (P1.2). Compared to (P1.2), the subproblem (P1.1) has a far lower computational complexity. The subproblem (P1.2) can be solved by cyclically optimizing the variables $\alpha$, $\boldsymbol{\beta}$, $\textbf{x}$, $\mathcal{V}$, and $\textbf{w}$. In the cyclical optimization, the problem (P1.5) dominates the four-step optimization. That is because the computational complexities of Steps 1), 2), and 4) are negligible owing to the closed expressions \eqref{Eq1} $\sim$ \eqref{Eq8}. According to the interior-point method (IPM),  the computational complexity of the problem (P1.5) is given by
\begin{align}
C_\text{p1.5} =&  \frac{\sqrt{3K(L+1)}}{\varepsilon} \Big[ 8 n_1 K (L+1)^3 + 4 n_1^2 K (L+1)^2 \nonumber\\
 &\quad\quad\quad\quad\quad\quad\quad\quad\quad\quad  + 4 (n_1^2 +  n_1) K (L+1) \Big], \nonumber
\end{align}
where $\varepsilon$ is the iteration accuracy with $n_1 = \mathcal{O}\{4 K (L+1)^2 \}$. Additionally, the computational complexity to recover the rank-one solution $\textbf{v}_{k}$
is negligibly small. Therefore, the total complexity of the problem (P1) is approximated as
\begin{align}
C_\text{p1} &\approx  M_\text{ite,1} C_\text{p1.5}, \nonumber
\end{align}
where $M_\text{ite,1}$ denotes the iteration number for the cyclical optimization.
\section{WTC Maximization for Binary Offloading}
\subsection{Optimization Scheme}
This section focuses on the optimization of the problem (P2) for the \emph{Binary Offloading} strategy. By comparing (P2) with (P1), it is seen that the differences between the two problems lie in the objective function and the constraints C7 and C8. In the \emph{Binary Offloading} strategy, the problem (P2) is rewritten as
\begin{subequations}
\begin{align}
(\text{P2.1}) \underset{\mathcal{V}, \mathcal{T}, \textbf{x}, \textbf{w}, \Xi} \max \quad & \omega_\text{r} t_\text{c} R_\text{r} + \sum_{k=1}^K  \omega_{k} \left[ \xi_k t_{\text{c}} R_{\text{c},k} + (1-\xi_k) t_{\text{loc}, k} R_{\text{loc}, k} \right], \nonumber \\
s.t. \quad
 & \text{C1} \sim \text{C8}. \nonumber
\end{align}
\end{subequations}
The problem (\text{P2.1}) is more challenging than (\text{P1.1}), due to the selection of integer variables $\xi_k$. To maximize the system WTC, we consider the following problem.
\begin{subequations}
\begin{align}
(\text{P2.2}) \quad \underset{\mathcal{V}, \textbf{x}, \Xi} \max \quad & \sum_{k=1}^K  \omega_{k} \xi_k  R_{\text{c},k}, \nonumber \\
s.t. \quad
 & \text{C1}, \text{C2}, \text{C7}, \text{C8}. \nonumber
\end{align}
\end{subequations}
By the similar method to solving the problem (\text{P1.2}), the problem (\text{P2.1}) can be addressed. Firstly, the objective function is transformed into
\begin{align}
&  \sum_{k=1}^K \omega_{k} \xi_k R_{\text{c},k}
=\underset{\alpha_{k} \geq 0,  \boldsymbol{\beta}_{k}} \max ~  \sum_{k=1}^K  \omega_{k} \xi_k \left[ \log \left( 1 + \alpha_{k} \right) - \alpha_{k} \right]   \nonumber\\
 & \quad\quad\quad +  2 \sqrt{\omega_{k}\xi_k ( 1 + \alpha_{k}) }  \text{Re} \{ \boldsymbol{\beta}_{k}^H \mathrm{A}_{\text{c},k} \}  -    \boldsymbol{\beta}_{k}^H \mathrm{B}_\text{c}  \boldsymbol{\beta}_{k}, \nonumber
\end{align}
Then, an AO procedure is performed to cyclically optimize the variables $\mathcal{V}$, $\Xi$, $\textbf{x}$, $\alpha$ and $\boldsymbol{\beta}$, including three steps: 1) optimizing $\alpha$ and $\boldsymbol{\beta}$; 2) optimizing $\textbf{x}$; 3) optimizing $\mathcal{V}$ and $\Xi$. The first two steps are similar to Section III-B and not repeated. In the following, we focus on jointly optimizing $\mathcal{V}$ and $\Xi$. \par
Similar to the problem (\text{P1.5}), the optimization of $\mathcal{V}$ and $\Xi$ is expressed as
\begin{align}
(\text{P2.3}) \quad \underset{\hat{\textbf{V}}_{k}, \Xi} \max \quad & - \sum_{k'=1}^K \text{Tr} \left(|{\beta}_\text{r}|^2 \textbf{T}_{\text{r}, k'} ^H \textbf{W}  \textbf{T}_{\text{r}, k'}  \textbf{V}_{k'}    \right)   \nonumber\\
&+ \sum_{k=1}^K  \Bigg[ \sqrt{ \omega_{k} \xi_k ( 1 + \alpha_{k}) }   \text{Tr} (\mathbf{\Omega}_{k}   \hat{\textbf{V}}_{k})  \nonumber\\
 &   -    \sum_{k'=1}^K \text{Tr} \left( \boldsymbol{\beta}_{k} \boldsymbol{\beta}_{k}^H {\textbf{T}}_{\text{c}, k'} \textbf{V}_{k'} {\textbf{T}}_{\text{c}, k'}^H \right) \Bigg], \nonumber\\
s.t. \quad
 & \text{C13}: \left[ \hat{\textbf{V}}_{k} \right]_{l,l}\leq 1, ~l \in \mathcal{L}, k \in \mathcal{K}, \nonumber\\
 & \text{C14}: \left[  \hat{\textbf{V}}_{k} \right]_{L+1,L+1} = 1, k \in \mathcal{K},  \nonumber\\
 & \text{C15}:  \hat{\textbf{V}}_{k} \succeq \textbf{0}, \nonumber\\
 & \text{C16}: \text{rank}\left(  \hat{\textbf{V}}_{k} \right) = 1, \nonumber\\
 & \text{C7}, \text{C8}. \nonumber
\end{align}
For the objective function of the problem (\text{P2.3}), it can be rewritten as
\begin{align}
&\sum_{k=1}^K - \text{Tr} \left(|{\beta}_\text{r}|^2 \textbf{T}_{\text{r}, k} ^H \textbf{W}  \textbf{T}_{\text{r}, k} \textbf{V}_{k}    \right)   + \sqrt{ \omega_{k} \xi_k ( 1 + \alpha_{k}) }   \text{Tr} (\mathbf{\Omega}_{k}  \hat{\textbf{V}}_{k})  \nonumber\\
 & \quad\quad\quad\quad\quad\quad\quad\quad\quad\quad   -    \sum_{k'=1}^K \text{Tr} \left( \boldsymbol{\beta}_{k'} \boldsymbol{\beta}_{k'}^H {\textbf{T}}_{\text{c}, k}  \textbf{V}_{k} {\textbf{T}}_{\text{c}, k}^H \right).  \nonumber
\end{align}
In the constraints of the problem (\text{P2.3}), $K$ optimization variables $ \hat{\textbf{V}}_{k}$ are mutually independent. Therefore, we consider the following problem.
\begin{align}
(\text{P2.4}) \quad \underset{\hat{\textbf{V}}_{k}} \max \quad & - \text{Tr} \left(|{\beta}_\text{r}|^2 \textbf{T}_{\text{r}, k} ^H \textbf{W}  \textbf{T}_{\text{r}, k}  \textbf{V}_{k}    \right) + \sqrt{ \omega_{k} ( 1 +   \alpha_{k}) }    \nonumber\\
 &   \times \text{Tr} (\mathbf{\Omega}_{k}   \hat{\textbf{V}}_{k})  -    \sum_{k'=1}^K \text{Tr} \left( \boldsymbol{\beta}_{k'} \boldsymbol{\beta}_{k'}^H {\textbf{T}}_{\text{c}, k} \textbf{V}_{k} {\textbf{T}}_{\text{c}, k}^H \right) \nonumber\\
s.t. \quad
 & \text{C7}, \text{C8}, \text{C13} \sim \text{C16}. \nonumber
\end{align}
Dropping the constraint C16, the problem (\text{P2.4}) is convex and easy to solve directly. Let  $\Gamma_{k}$ denote the maximum value of the objective function of the problem (\text{P2.4}) for the $k$-th UE. Then, the problem (P2.3) for given $\Gamma_{k}$ is simplified as a weighted bipartite matching problem as follows.
\begin{align}
(\text{P2.5}) \quad \underset{\Xi} \max ~ & \sum_{k = 1 }^K \xi_{k} \Gamma_{k}, \nonumber\\
s.t. \quad
 & \text{C7}, \text{C8}. \nonumber
\end{align}
It is not difficult to find that $\xi_{k} = 1$ if $\Gamma_{k}$ is no less than their median.
\subsection{Complexity Analysis}
\begin{algorithm}[t]
\caption{Overall algorithm for the problem (P2)}
\begin{algorithmic}[1]
\State \textbf{Initialization}: Set $m$ = 0, $\varepsilon$, $\textbf{x}$, $\textbf{w}$, $\mathcal{V}$, $f_\text{opt}^{(m)}$.
\State \textbf{repeat}
\State \quad Set $m = m + 1$.
\State \quad Compute $\alpha_\text{r}^{(m)}$, $\alpha_k^{(m)}$, $\beta_\text{r}^{(m)}$ and $\boldsymbol{\beta}_k^{(m)}$ by \eqref{Eq1} $\sim$ \eqref{Eq4}. \nonumber
\State \quad  Find $\textbf{x}^{(m)}$ by \eqref{Eq5} and \eqref{Eq6}. \nonumber
\State \quad  Jointly optimize (\text{P2.4}) and (\text{P2.5}), followed by rank-one recovery, to obtain $\mathcal{V}^{(m)}$ and $\Xi^{(m)}$. \nonumber
\State \quad  Find $\textbf{w}^{(m)}$ by \eqref{Eq7} and \eqref{Eq8}. \nonumber
\State \quad  Compute $f_\text{opt}^{(m)} = f(\mathcal{V}^{(m)}, \Xi^{(m)}, \textbf{x}^{(m)}, \textbf{w}^{(m)}, \alpha^{(m)}, \boldsymbol{\beta}^{(m)}) $. \nonumber
\State \textbf{until} $\frac{f_\text{opt}^{(m)} - f_\text{opt}^{(m-1)}}{f_\text{opt}^{(m)}} < \varepsilon$.
\State Employ LP to solve (\text{P2.1}) to obtain the optimal $C_{\text{bo}}^{\ast}$.
\State \textbf{return} $C_{\text{bo}}^{\ast}$.
\end{algorithmic}
\end{algorithm}
\begin{figure*}[!ht]
  \centering
  \subfigure[Partial Offloading]{\label{Fig3a}
            \includegraphics[width=3.3in]{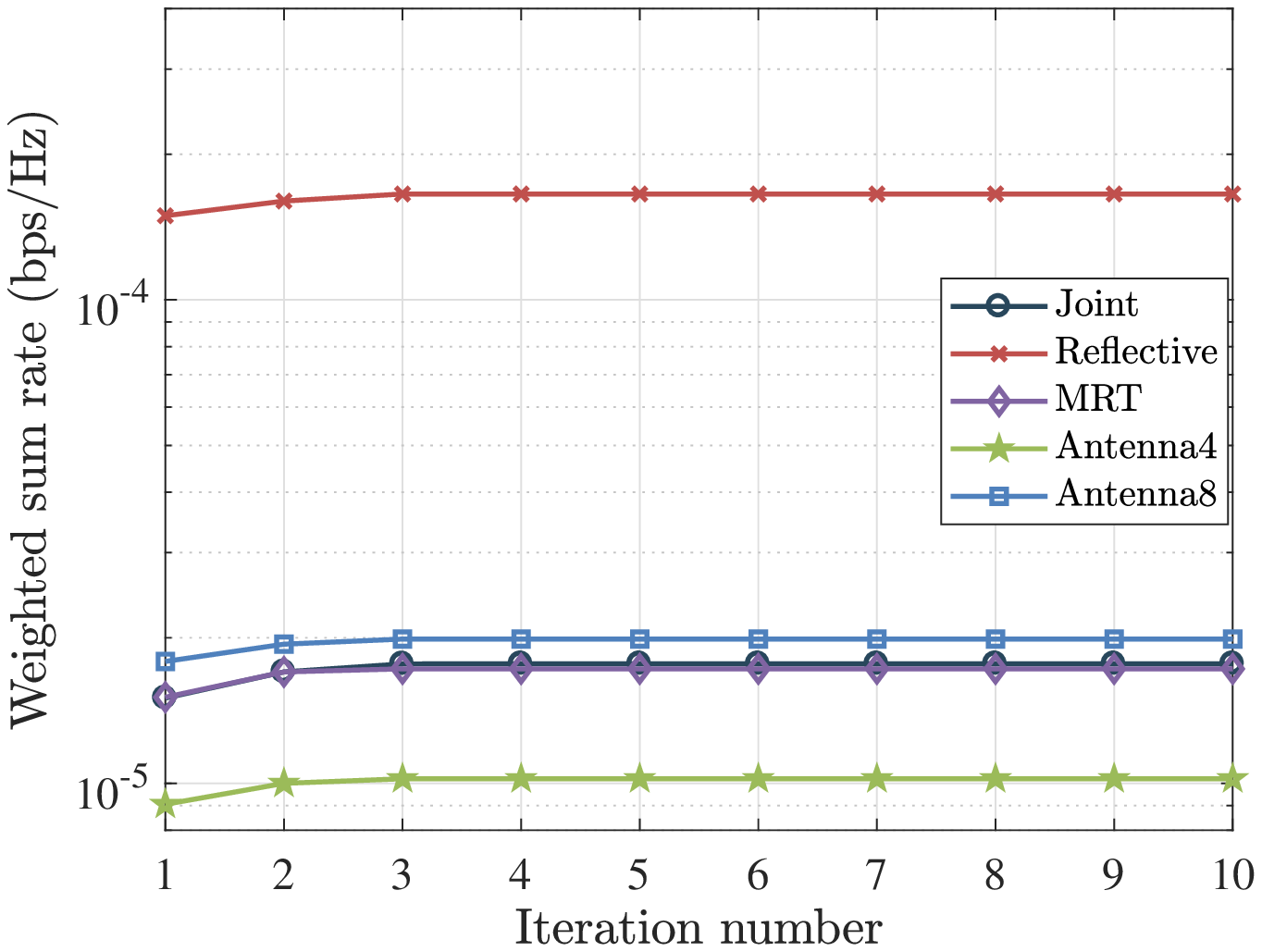}}
  \subfigure[Binary Offloading]{\label{Fig3b}
            \includegraphics[width=3.3in]{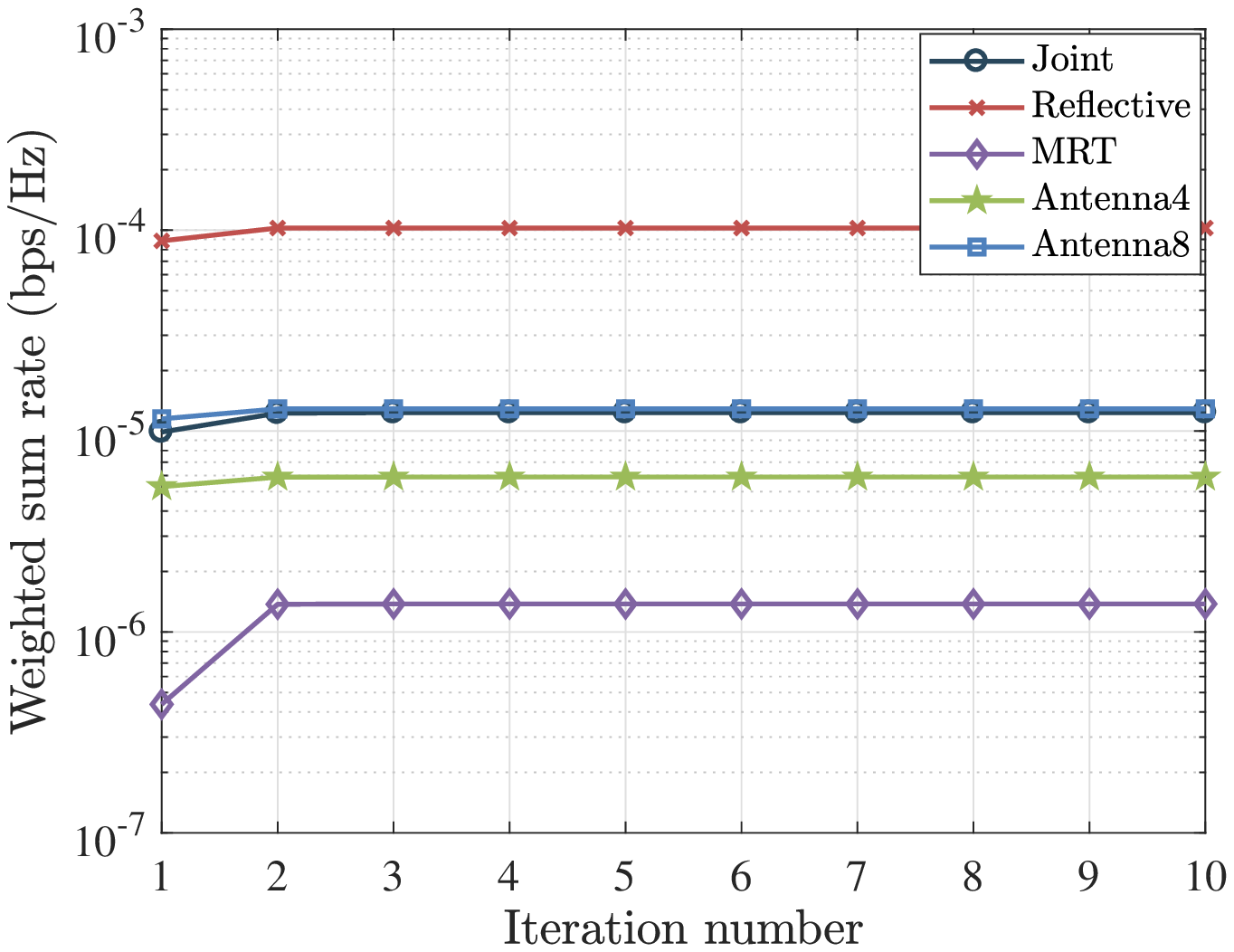}}
  \caption{Convergence behavior of all optimization schemes for the considered IRS backscatter enabled ISCC system in a random observation, including the proposed optimization schemes and the benchmark schemes. }
  \label{Fig3}
\end{figure*}
The overall algorithm for the problem (P2) is given in Algorithm 2. The problem (P2) can be solved by its two subproblems (P2.1) and (P2.2). The subproblem (P2.1) has a far lower computational complexity than (P2.2). The subproblem (P2.2) can be solved by cyclically optimizing the variables $\alpha$, $\boldsymbol{\beta}$, $\textbf{x}$, $\mathcal{V}$ and $\Xi$. In the cyclical optimization, the problem (P2.4) dominates the computational complexity of the subproblem (P2.2). According to the IPM,  the computational complexity of the problem (P2.2) is given by
\begin{align}
C_\text{p2.4} =&  \frac{\sqrt{3(L+1)}}{\varepsilon} \Big[ 8 n_2  (L+1)^3 + 4 n_2^2  (L+1)^2 \nonumber\\
 &\quad\quad\quad\quad\quad\quad\quad\quad\quad\quad  + 4 (n_2^2 +  n_2)  (L+1) \Big], \nonumber
\end{align}
where $n_2 = \mathcal{O}\{4  (L+1)^2 \}$. Therefore, the total complexity of the problem (P2) is approximated as
\begin{align}
C_\text{p2} &\approx  M_\text{ite,2} K C_\text{p2.4}, \nonumber
\end{align}
where $M_\text{ite,2}$ denotes the iteration number for the cyclical optimization.
\begin{table}[]
\scriptsize
\begin{center}
\caption{Simulation parameters.}\label{T1}~~\\
\begin{tabular}{c|c|c}
\hline
Notation        &  Description                    & Value       \\ \hline
\hline
$\kappa$        &  Rician factor                  & 3           \\ \hline
$d_0$           &  Reference distance             & 1m          \\ \hline
$d_\text{r}$           &  Distance from BS to target     & 50 m            \\ \hline
$d_\text{u}$           &  Average distance from BS to UEs     & 50 m        \\ \hline
$N_\text{t}$             &  Number of transmitting antennas at BS          & 4           \\ \hline
$N_\text{r}$             &  Number of radar-receiving antennas at BS             & 4           \\ \hline
$N_\text{c}$             &  Number of information-receiving antennas at BS             & 4           \\ \hline
$N_\text{a}$             &  Number of active antennas at each UE in \texttt{Antenna}          & 4  \text{or} 8         \\ \hline
$K$                      &  Number of UEs                                                           & 4           \\ \hline
$L$             &  Number of IRS elements          & 64          \\ \hline
$\sigma^2$      &  Noise variance                 & -10 dBm     \\ \hline
$\omega_{k}$    &  Weighting factor               & 1           \\ \hline
$B$             &  Bandwidth                      & 1   GHZ        \\ \hline
$P$             &  Total power of BS             &3 dBW        \\ \hline
$P_\text{a}$    &  Transmit power of each UE  in \texttt{Antenna}              &-6 dBm        \\ \hline
$f_k$           &  CPU's frequency of each UE        & 1 GHz       \\ \hline
$c_k$           &  Cycle number of  processor's chip at each UE                      & $10^5$          \\ \hline
$\varepsilon_k$   &Energy consumption coefficient of processor's chip & 10 $\text{W/GHz}^3$  \\ \hline
$E_k^\text{th}$             & Energy threshold value of each UE                  & $0.8 E_k$         \\ \hline
$T$                         &Time block                                         & 100 s         \\ \hline
\end{tabular}
\end{center}
\end{table}
\begin{figure*}[!ht]
  \centering
  \subfigure[Partial Offloading]{\label{Fig4a}
            \includegraphics[width=3.3in]{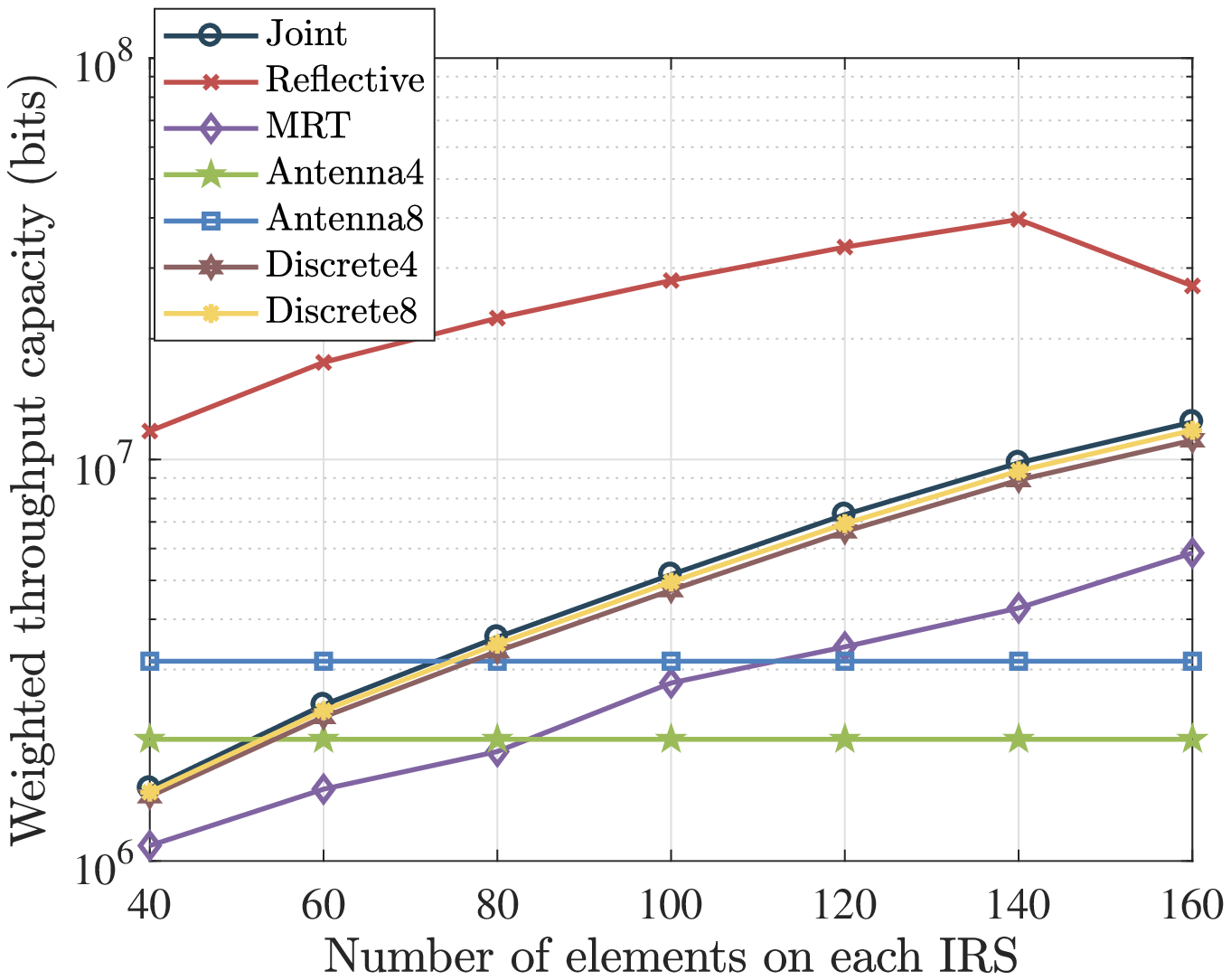}}
  \subfigure[Binary Offloading]{\label{Fig4b}
            \includegraphics[width=3.3in]{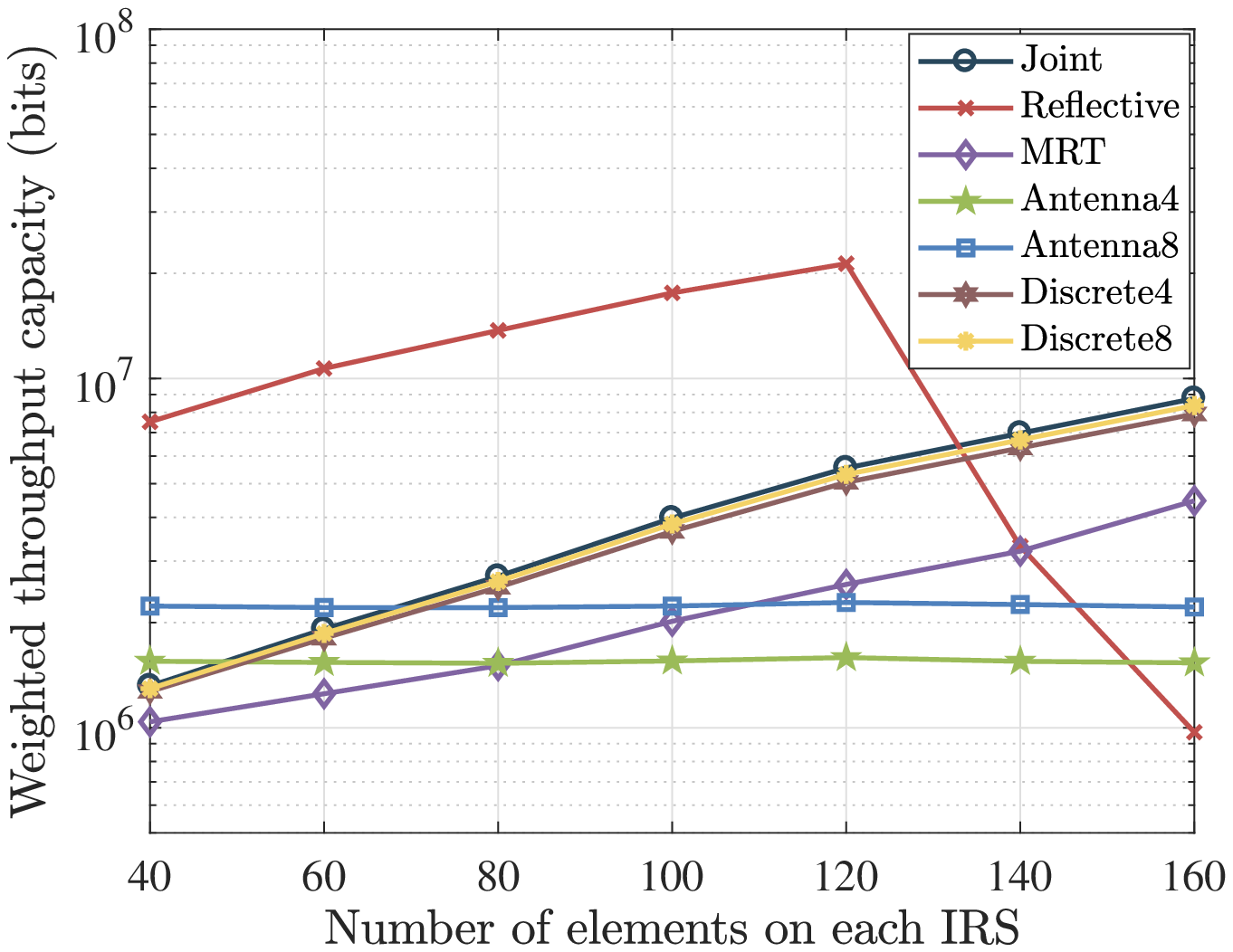}}
  \caption{The relationship between the number of IRS elements and the achievable WTC for the considered IRS backscatter enabled ISCC system. }
  \label{Fig4}
\end{figure*}
\section{Numerical Results}
This section will present simulation results to show the achievable performance of the proposed optimization schemes for the considered IRS backscatter enabled ISCC system.
In addition to the proposed optimization schemes, several counterparts are simulated for comparison.
\begin{itemize}
 \item \texttt{Joint}: This legend denotes the proposed optimization scheme for \emph{Partial Offloading} or \emph{Binary Offloading} in the considered IRS backscatter enabled ISCC system. In \texttt{Joint}, the transmitting beamforming at the BS, the passive beamforming at all the IRSs, the radar receiving beamforming, the time of data computing and communication (and the integer variables) are jointly optimized, as specified in Section III and IV.
 \item \texttt{MRT}: This legend denotes the simplified optimization scheme for \emph{Partial Offloading} or \emph{Binary Offloading} in the considered IRS backscatter enabled ISCC system. In \texttt{MRT}, the normalized receiving beamforming vector \textbf{w} and the transmitting beamforming vector \textbf{x} are respectively set as $\textbf{w} = \frac{ \textbf{1}}{N_\text{r}}$ and $\textbf{x} = \frac{\sqrt{P} \textbf{w}^H \mathbf{\Lambda}_\text{r}(\theta) }{|\textbf{w}^H \mathbf{\Lambda}_\text{r}(\theta)| }$, so as to track the radar target specially.
 \item \texttt{Antenna}: This legend denotes a benchmark scheme of employing active antennas at each UE instead of IRS, with '4' and '8' being the number of antennas at each UE. Additionally, similar to \texttt{MRT}, the normalized receiving beamforming vector \textbf{w} and the transmitting beamforming vector \textbf{x} are respectively set as $\textbf{w} = \frac{ \textbf{1}}{N_\text{r}}$ and $\textbf{x} = \frac{\sqrt{P} \textbf{w}^H \mathbf{\Lambda}_\text{r}(\theta) }{|\textbf{w}^H \mathbf{\Lambda}_\text{r}(\theta)| }$.
 \item \texttt{Reflective}: This legend denotes a benchmark scheme, in which the normalized receiving beamforming vector \textbf{w} and the transmitting beamforming vector \textbf{x} are respectively set as $\textbf{w} = \frac{ \textbf{1}}{N_\text{r}}$ and $\textbf{x} = \frac{\sqrt{P} \textbf{w}^H \mathbf{\Lambda}_\text{r}(\theta) }{|\textbf{w}^H \mathbf{\Lambda}_\text{r}(\theta)| }$, while the reflection power is equally distributed at each element of IRS. What is more, the total reflection power is the same as the transmit power at each UE in \texttt{Antenna}.
 \item \texttt{Discrete}: This legend represents the discrete optimization scheme corresponding to \texttt{Joint}, where '4' and '8' denote the number of discrete values for the passive beamforming at each IRS.
\end{itemize}
\begin{figure*}[!ht]
  \centering
  \subfigure[Partial Offloading]{\label{Fig5a}
            \includegraphics[width=3.3in]{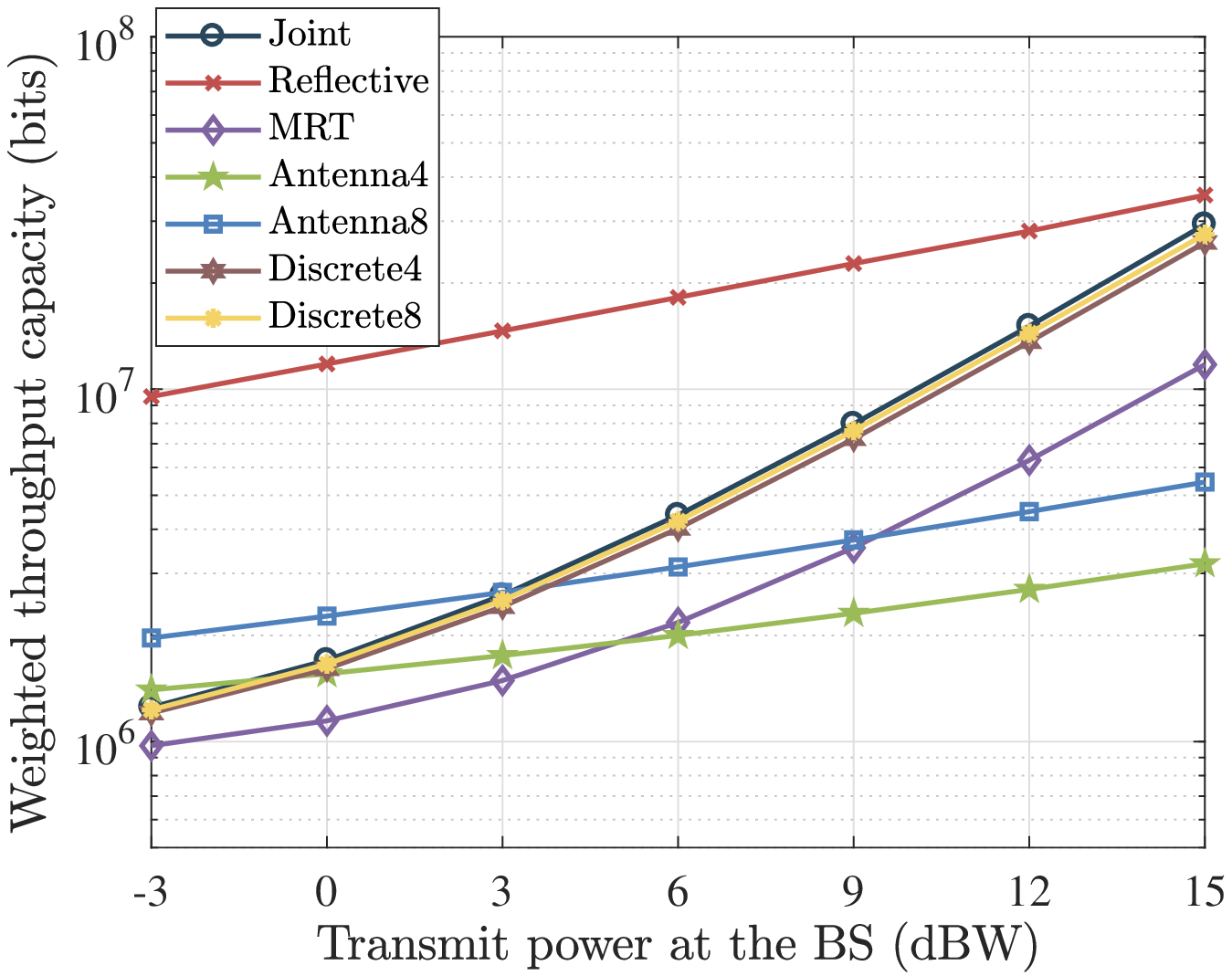}}
  \subfigure[Binary Offloading]{\label{Fig5b}
            \includegraphics[width=3.3in]{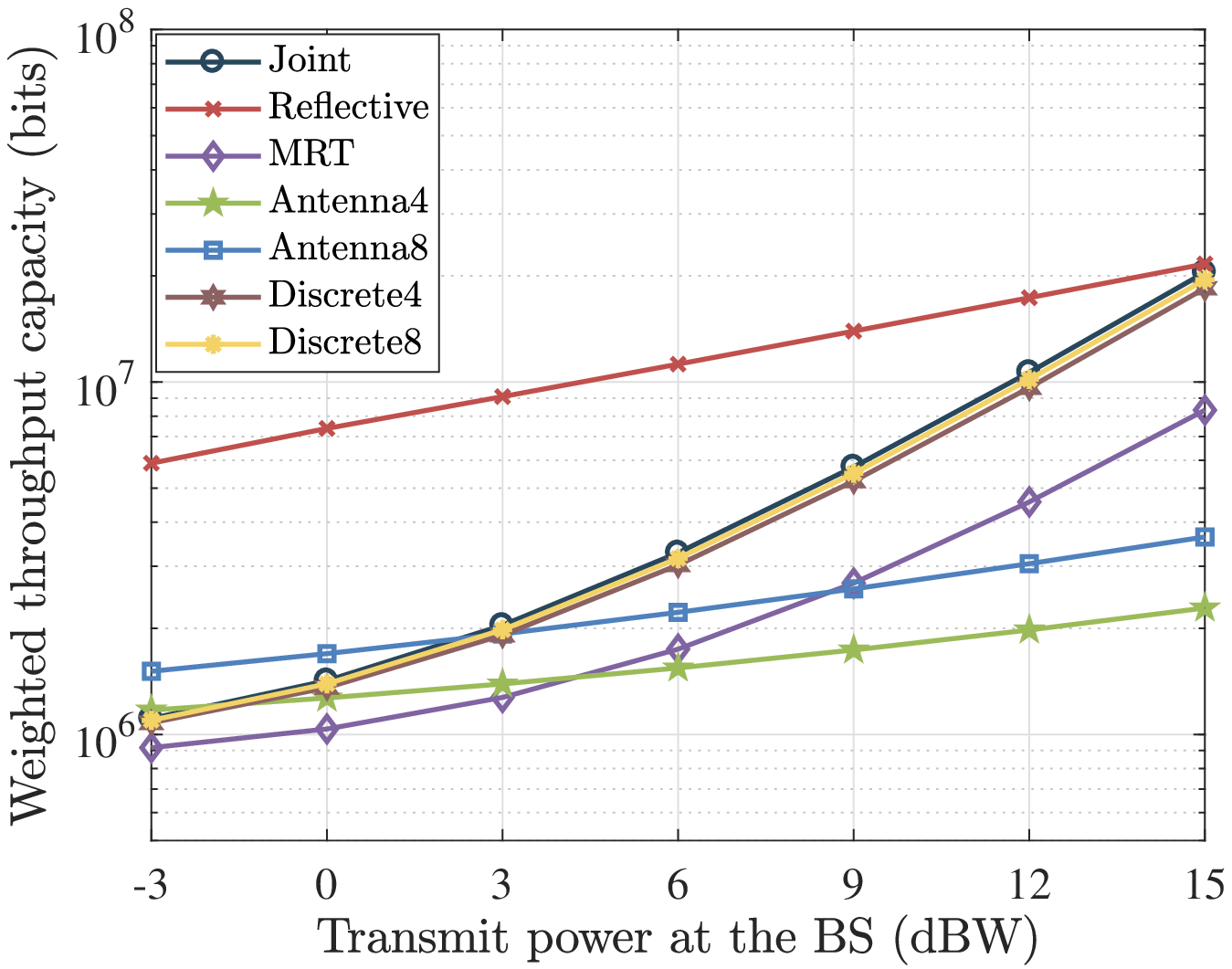}}
  \caption{The relationship between the total transmit power at the BS and the achievable WTC for the considered IRS backscatter enabled ISCC system. }
  \label{Fig5}
\end{figure*}
\begin{figure*}[!ht]
  \centering
  \subfigure[Partial Offloading]{\label{Fig6a}
            \includegraphics[width=3.3in]{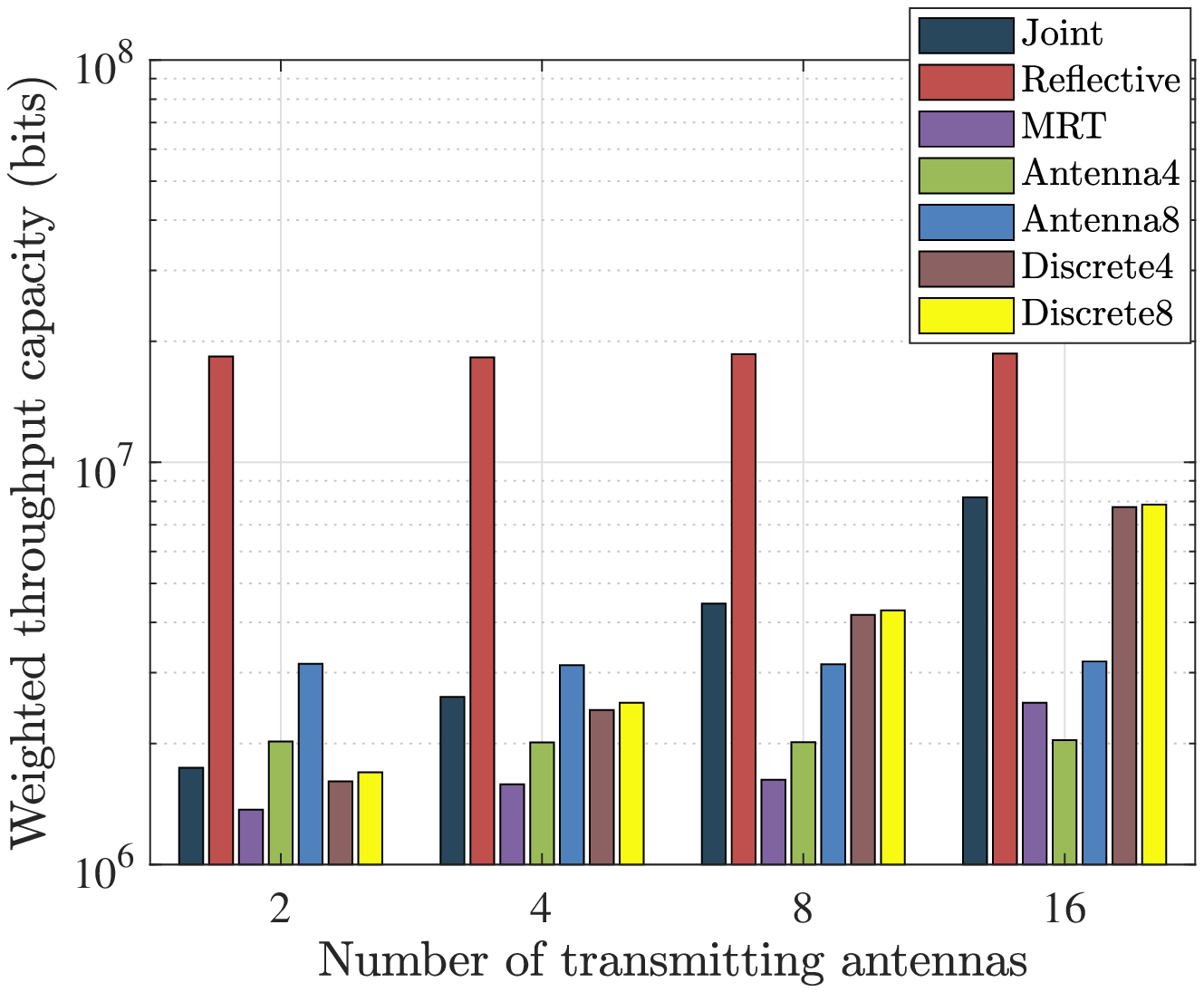}}
  \subfigure[Binary Offloading]{\label{Fig6b}
            \includegraphics[width=3.3in]{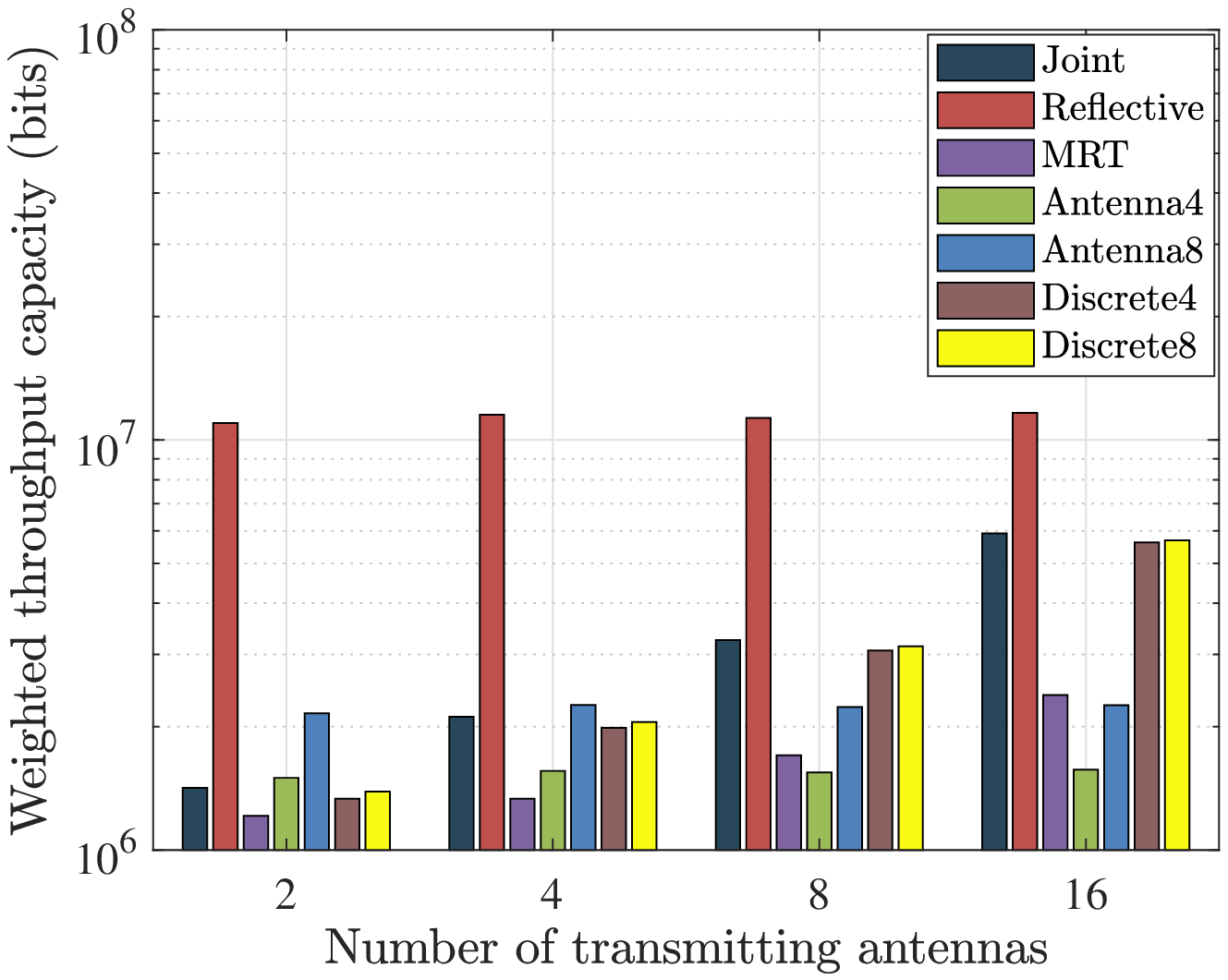}}
  \caption{The relationship between the number of transmitting antennas at the BS and the achievable WTC for the considered IRS backscatter enabled ISCC system. }
  \label{Fig6}
\end{figure*}
In simulations, it is assumed that all communication channels arriving at or departing from the IRSs follow Rician distribution with the same Rician factor $\kappa$. For the transmitting and receiving antennas at the DFRC BS, uniform linear arrays (ULAs) are employed with half-wavelength antenna spacing. For all communication and radar channels, the path-loss coefficient is modelled as $\text{PL} = \text{PL}_0 - 25 \lg \left(d/{d_0} \right) $ dB, where $d$ and $d_0$ represent the transmission distance and the reference distance, respectively, with $\text{PL}_0$ = -30 dB. The distance from the BS to each UE is generated randomly from $[d_\text{u} - 10, d_\text{u} + 10]$, where $d_\text{u}$ represents the average distance from the BS to the UEs. Additionally, each UE in \texttt{Antenna} is assumed to has the same power consumption as \texttt{Joint} for fair comparison. In TABLE \ref{T1}, most involved parameters for the simulations for the considered IRS backscatter enabled ISCC system are listed. Unless stated otherwise, the involved parameters in the simulations generally refer to the given constant values.  \par
Fig. \ref{Fig3} depicts the convergence behavior of all optimization schemes for the considered IRS backscatter enabled ISCC system in a random observation, including the proposed optimization scheme and the benchmark schemes. From Figs. \ref{Fig3a} and \ref{Fig3b}, it is seen that the AO optimization schemes of \texttt{Joint}, \texttt{Reflective}, \texttt{MRT}, \texttt{Antenna4} and \texttt{Antenna8} converge very quickly for both \emph{Partial Offloading} and \emph{Binary Offloading}. According to the results, it is found that the number of iterations is small. \par
\begin{figure*}[!ht]
  \centering
  \subfigure[Partial Offloading]{\label{Fig7a}
            \includegraphics[width=3.3in]{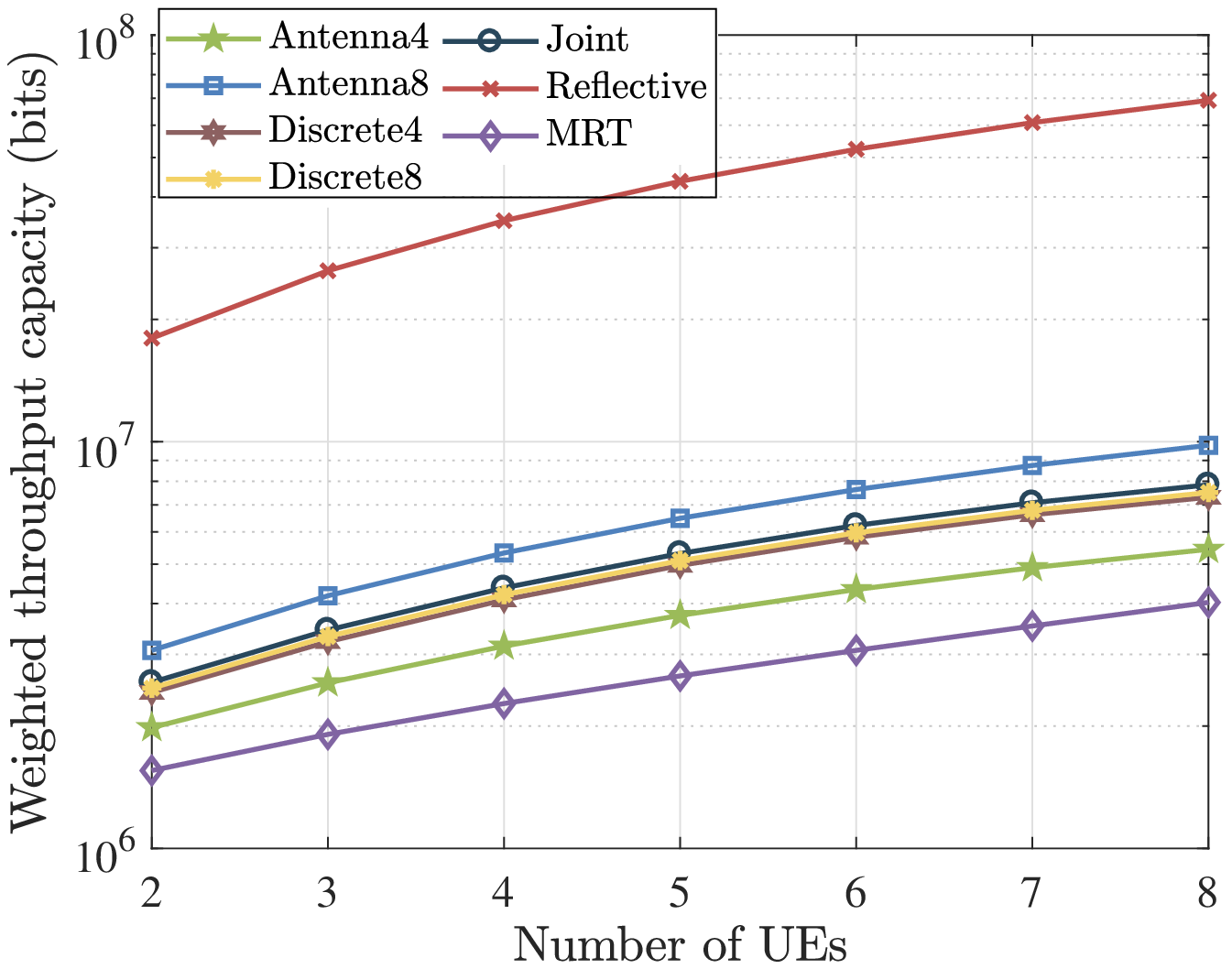}}
  \subfigure[Binary Offloading]{\label{Fig7b}
            \includegraphics[width=3.3in]{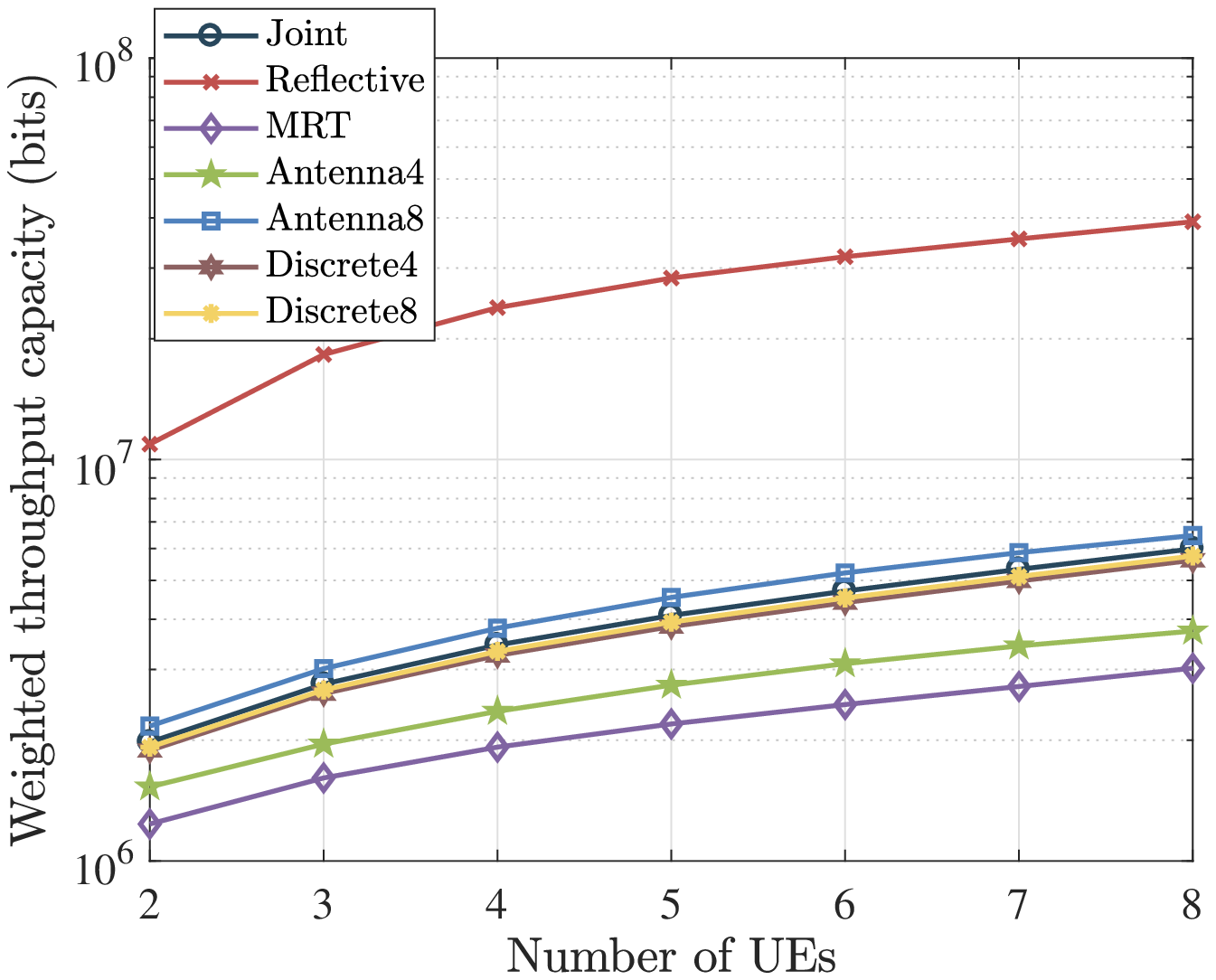}}
  \caption{The relationship between the number of UEs and the achievable WTC for the considered IRS backscatter enabled ISCC system.}
  \label{Fig7}
\end{figure*}
Fig. \ref{Fig4} shows the impact of the number of IRS elements on the achievable WTC for the considered IRS backscatter enabled ISCC system. It is observed that an increase in the number of IRS elements contributes to improving the achievable WTC for the schemes of \texttt{Joint}, \texttt{MRT}, \texttt{Discrete4} and \texttt{Discrete8}. That is because more elements coated on each IRS enable more signal power reception as well as a higher degree of spatial freedom. However, the WTC for the schemes of \texttt{Reflective} first goes up and then falls as the number of IRS elements increases. The main reason is that equally distributed power at each element of IRS results in a smaller element reflection power. Moreover, the high degree of spatial freedom is not fully exploited due to equal power allocation among all elements. On the other hand, the schemes of \texttt{Joint}, \texttt{MRT}, \texttt{Discrete4} and \texttt{Discrete8} can achieve a higher WTC than \texttt{Antenna4} and \texttt{Antenna8}, which indicates that IRS backscatter may be a substitute for active antennas to realize UT to some extent.\par
Fig. \ref{Fig5} presents the impact of the total transmit power at the BS on the achievable WTC for the considered IRS backscatter enabled ISCC system, where the total reflection or transmit power in \texttt{Reflective}, \texttt{Antenna4} and \texttt{Antenna8} ranges from -9 dBm to -3 dBm, with the step length being 1 dBm. As the total transmit power at the BS increases, the achievable system WTC increases.
Fig. \ref{Fig6} shows how the achievable WTC depends on the number of transmitting antennas at the BS for the considered IRS backscatter enabled ISCC system. From Fig. \ref{Fig6}, it is found that an increase in the number of transmitting antennas at the BS is beneficial to the improvement of WTC for the schemes of \texttt{Joint}, \texttt{MRT}, \texttt{Discrete4} and \texttt{Discrete8}. For the schemes of \texttt{Reflective}, \texttt{Antenna4} and \texttt{Antenna8}, however, the number of transmitting antennas at the BS has little effect. Fig. \ref{Fig7} shows how the achievable WTC is affected by the number of UEs for the considered IRS backscatter enabled ISCC system. In Fig. \ref{Fig7}, the number of information-receiving antennas at the BS is set as eight so as to support eight data streams from eight UEs at the same time. These results of Figs. \ref{Fig6} and \ref{Fig7} imply that high degree of spatial freedom owing to more antennas or UEs can facilitate communication performance of system.\par
From Figs. \ref{Fig3} $\sim$ \ref{Fig7}, we observe that the scheme of \texttt{Joint} outperforms \texttt{MRT} and is slightly better than \texttt{Discrete4} and \texttt{Discrete8}. These results confirm the superiority of the proposed scheme and the feasibility of discrete passive beamforming at each IRS. By comparing the scheme of \texttt{Reflective} with \texttt{Antenna}, it is found that IRS backscatter is often superior to active antennas for the same transmitting power at each UE. That is because IRS has a higher degree of spatial freedom. This result demonstrates the feasibility and the practicability of IRS backscatter. On the other hand, the achievable WTC in \emph{Partial Offloading} is always higher than \emph{Binary Offloading}. That is because all UEs can only select either local computation or data offloading in \emph{Binary Offloading}. \par
\section{Conclusions}
This paper proposed the concept of IRS backscatter enabled RFCF-UT integrally. Moreover, an illustrative example of an IRS backscatter enabled
ISCC system was given to explain how the proposed technique works. Based on the established model of ISCC system, two optimization problems were formulated and then addressed by using LP, FP, IP and OA to jointly optimize the transmitting beamforming at the BS, the passive beamforming at all the IRSs, the radar receiving beamforming, the time of data computing and communication (and the integer variables). According to the simulation results, it was verified that: 1) the proposed optimization schemes for the WTC maximization problems are feasible and have a superiority to the simplified optimization schemes of \texttt{MRT}; 2) IRS backscatter can achieve a good communication performance as active antennas in many cases; 3) the communication performance achieved by discrete passive beamforming at each IRS is in close proximity to that of the corresponding continuous passive beamforming for the considered ISCC system; 4) the proposed paradigm of IRS backscatter enabled RFCF-UT is validated.
\ifCLASSOPTIONcaptionsoff
  \newpage
\fi
%

%
%
%




\end{document}